# The Golden Eras of Graphene Science and Technology: Bibliographic Evidences From Journal and Patent Publications

Ai Linh Nguyen,[1,4] Wenyuan Liu,[1,4] Khiam Aik Khor,[2] Andrea Nanetti,[3] Siew Ann Cheong[1,4]

[1]Division of Physics and Applied Physics, School of Physical and Mathematical Sciences, Nanyang Technological University, 21 Nanyang Link, Singapore 637371
[2]School of Mechanical & Aerospace Engineering, Nanyang Technological University, 50 Nanyang Avenue, Singapore 639798
[3]School of Art, Design and Media, Nanyang Technological University, 81 Nanyang Dr, Singapore 637458
[4]Complexity Institute, Nanyang Technological University, 50 Nanyang Avenue, Singapore 639798

**Abstract**

Today's scientific research is an expensive enterprise funded largely by taxpayers' and corporate groups' monies. It is a critical part in the competition between nations, and all nations want to discover fields of research that promise to create future industries, and dominate these by building up scientific and technological expertise early. However, our understanding of the value chain going from science to technology is still in a relatively infant stage, and the conversion of scientific leadership into market dominance remains very much an alchemy rather than a science. In this paper, we analyze bibliometric records of scientific journal publications and patents related to graphene, at the aggregate level as well as on the temporal and spatial dimensions. We find the present leaders of graphene science and technology emerged rather late in the race, after the initial scientific leaders lost their footings. More importantly, notwithstanding the amount of funding already committed, we find evidences that suggest the 'Golden Eras' of graphene science and technology were in 2010 and 2012 respectively, in spite of the continued growth of journal and patent publications in this area.

## 1. Introduction

The 2010 Nobel Prize in Physics was awarded to Andre Geim and Konstantin Novoselov, both from the University of Manchester in UK then ("The Nobel Prize in Physics 2010," n.d.), "for groundbreaking experiments regarding the two-dimensional material graphene" (Novoselov et al., 2004). Since their 2004 discovery, there has been a flurry of research activities on the material; and given superlative claims on possible future technologies based on it, attracted huge amounts of grant money from public and private sources.

A consortium led by Jari Kinaret was funded to the tune of 1 billion EUR to set up the Graphene Flagship ("Graphene Flagship | Graphene Flagship," n.d.) in 2013, to bring together researchers in academia and industry for Europe to spearhead commercial applications of graphene within 10 years. Elsewhere in Asia, the Korean government also created the Korean Graphene Hub, an independent government institute originally intended to become operational in 2018, with an annual budget of 200-300 million USD ("South Korea Funds Graphene Commercialization – NextBigFuture.com," n.d.). The creation of this Korean Graphene Hub (to focus on the fundamental sciences of graphene and related 2D materials) follows 64.6 million USD of funding by South Korea on graphene projects between 2006 and 2011. During the same period (2006-2010), the US spent 74.6 million USD on 168 graphene projects, while EU spent 68.8 million EUR on 47 projects ("CORDIS | European Commission," n.d.; "National Science & Technology Information Service(NTIS)," n.d.; "NSF - National Science Foundation," n.d.). In spite of its small geographical and economic size, Singapore has also invested heavily into graphene research. In fact, it has invested more per capita into the material than any other country, and boasts a dedicated and world-class Centre for Advanced 2D Materials ("Centre for Advanced 2D Materials," n.d.).

Shortly after graphene was discovered, many revolutionary technological applications were promised, both in scientific papers, (Avouris & Dimitrakopoulos, 2012; Palacios, 2011; Rao, Sood, Subrahmanyam, & Govindaraj, 2009) and also in popular presses ("Digital Trends," n.d.; "TECHNOLOGIST," 2014). Over the years, some of these touted applications did not live up to their promises, while many other potential applications were suggested. To date, the applications considered closest to commercialization, based on their economic efficiency and technological feasibility, include: (1) touch panels and organic light-emitting diodes (OLEDs), (2) batteries and supercapacitors, and (3) electro-magnetic shielding.

Now if all the above potential applications were realized, what would be the gross market value of graphene? According to a Statistica survey published in February 2019 [https://www.statista.com/statistics/972251/us-graphene-market-value/], the graphene market in the US in 2016 is worth 5.1 million USD, and forecasted to grow to 58.9 million USD by 2024. In China, which has 58 per cent of the world's graphene patents, the graphene market is worth more than 4 billion CNY (610.8 million USD) in 2016 according to the Global Graphene Industry Report 2017 ("China No 1 in world patent applications for graphene tech - Chinadaily.com.cn," n.d.), and expected to exceed 10 billion CNY in 2017. This same report estimated that the global graphene market will be worth 100 billion CNY by 2020, whereas in their report "Global Graphene Market Size 2017 By Type (Graphene Nanoplatelets, Graphene Oxide, and Others), by Application (Electronics, Composites, Energy, and Others), and by Region and Forecast 2018 to 2025", Adroit Market Research estimated that the global graphene market will be worth 221.4 million USD by 2025 ("Graphene Market Analysis by Size, Price, Demand, Growth | Industry Report 2019-2025," n.d.). The disparate estimates notwithstanding, it is clear that graphene technology will have significant commercial value.

Lured by the promises of commercial returns, many governments and corporations are investing heavily to get a piece of the graphene 'pie'. Unfortunately, our understanding of the RIE process (see Figure 1) is not yet at the level where science automatically and immediately becomes technology, which then becomes commercial products. Fundamentally, there is a value chain going from basic research (theory, synthesis & characterization) to applied research (device) to technology (patents) to product development to commercialization. To fully understand how a scientific idea on graphene ends up as a commercial product, we will need to collect and analyze many different data sets, and then more importantly, understand how advances and breakthroughs at one level of the value chain lead to advances and breakthroughs in the next level of the value chain. This is an ambitious undertaking, and is presently constrained by the many dispersed data sets that we need to bring together. Nevertheless, given the already large investment that has gone into graphene research, we would like to put ourselves in the shoes of a government or a corporation and ask whether these are all worth it, even before graphene as a material starts to generate serious commercial *returns on investment (ROI)*.

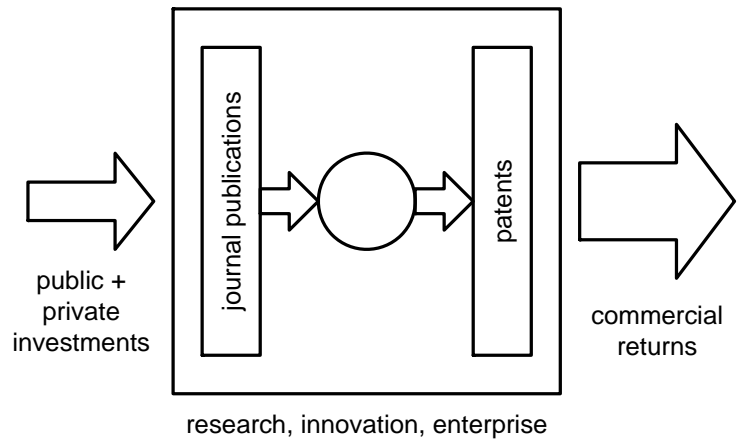

Figure 1. A schematic showing how public and private investments can generate commercial returns through the research, innovation, enterprise (RIE) process, shown as a large block. Within the RIE block, we show journal publications as the primary outcome of research, and patents as the primary product of innovation. The two are connected through a knowledge process that is still poorly understood.

To answer this question directly, we propose to measure the amounts of public and private investments (in dollar value), and thereafter also measure the commercial values of graphene products (in dollar value). If the latter is larger than the former, we will know that our investment is paying off. Alternatively, if the latter is still smaller than the former, but is increasing with time, then we hope that our investment will one day pay off. Unfortunately, it is difficult to directly measure the commercial value of graphene products, firstly because there is no systematic data on them (leading to the discrepancy between different estimates of graphene market values). Secondly, the current number of commercial products based on graphene is still small, and much of the graphene research has yet to bear fruits, so estimates based on these grossly underestimates the market value. We can of course wait ten more years, when graphene-based technology is more mature, to compute its commercial value over a wider range of products. However, until then it is highly likely that governments and corporations would continue pumping money into graphene science and technology, so if at that point in time we realized that graphene is not delivering on its initial promise, there would be no way to recover the investment. Therefore, it is important to somehow make an estimate of the ROI *now*, so that we can decide whether or not to follow up with ten more years of investment. Finally, companies can profit from an innovation or a product long after it was first introduced. In spite of early efforts by Nordhaus (Nordhaus, 1967), who introduced the first model of lifetime value by assuming a constant rate of return, and improvements by Matutes et al. (Matutes, Regibeau, & Rockett, 1996), who assumed the rate of return increases and then decreases, the *lifetime value* of an innovation or a product is difficult to estimate in general. In fact, Reitzig (Reitzig, 2003) tested different factors that are believed to affect patent value and found that even the patent lifetime itself is difficult to measure. To make matters worse, we further argue that it is also difficult to measure investments. While public investments are announced, and their dollar values reported, only part of the private investments are known through corporate announcements. Moreover, in the creation of scientific knowledge on graphene, it is entirely possible in some places for a university professor to devote his/her time and the time of his/her PhD students to graphene research. Such unfunded research adds to the body of scientific knowledge for graphene, but the dollar value of this investment in time cannot be easily computed.

Therefore, in this paper we use proxies to measure the input and output of RIE efforts related to graphene. For the input, in place of public or private investments we look at journal publications related to graphene. We argue that this is a more accurate measure of input than reported public/private investments, because the time invested by university professors and their PhD students can be measured on equal footing to research grants. To show that this contribution to the input is not negligible, we checked the 127,546 graphene journal publications, and found 20,539 (16.1%) that listed no information on funding. To be certain that these are not simple omissions, we check the list of authors against the funding statuses of the graphene papers. Here we assume that if an author is funded (from the acknowledgments of his/her papers), then all or nearly all of his/her papers should be funded, even if he/she did not acknowledge funding in a number of them. By this criterion, we realized that there are 186,460 funded authors who have at least one graphene paper acknowledging funding. In contrast, 11,000 unfunded authors produced 5,024 graphene papers. Therefore, we are talking about a silent investment in time that is about 4% of the total scientific knowledge produced. This proportion is not large, but certainly not small enough to be ignored. For the output, we propose to use patents as a proxy for the potential lifetime commercial value of graphene technology. For the purpose of measuring potential commercial value before the technology hits the markets, there are probably no better proxies.

In Section 2 of this paper, we will review the literature on journal and patent publications to justify our choice of proxies. Following this, we describe the data we used for our study, and how we analyze the data in Section 3. In Section 4, we organized our results into three main sections. In Section 4.1, we show the publications, their references, and their citations in the form of inflow-outflow diagrams for journal papers and for patents. We find here that the number of citations is smaller than the number of references, which suggests declining interest in graphene science and technology. We show this declining interest more clearly in Section 4.2, where we plot the yearly numbers of publications, references, and citations. By breaking the journal references up into those related to graphene, nanotube, 2D materials, and others, we see graphene papers making up the dominant component of the references of graphene papers between 2008 and 2015. Similarly, when we break the citing patent up into those related to graphene, nanotube, 2D materials, and others, we find graphene patents being the dominant component between 2009 and 2012. These suggest the existence of 'golden eras' in graphene science and technology during these periods. In Section 4.3, we considered the spatial/geographical dimension, to find China as the leader in graphene science and technology, followed some distance away by the US. We also look at the flow of citations between the top regions, and found that beyond the strong citations within regions, the strongest inter-region citations are by EU and WPO patents, citing mostly US patents. For journal papers, beyond citing themselves, the top regions cite equally strongly papers from US and China. We also show the rankings of the top regions in publications or patents, and how they evolve over time. In Section 5, we report more compelling evidence for the 'golden eras', by calculating the ratio of empirical average number of citations to the expected average number of citations. Finally, we conclude in Section 6.

## 2.    Literature Review

Let us start by justifying our choice of proxies for measuring the total investment and the total potential commercial value. On the input end, we chose the number of journal publications as the proxy for total investment (public, private, and time), since the average funding required to produce a scientific publication is more or less constant after adjusting for inflation. Even though Leydesdorff and Wagner (Leydesdorff & Wagner, 2009) concluded that the price per paper cannot be estimated, after analyzing the relations between research funding and research output in 34 countries/regions, there were also others who succeeded in estimating this cost of publication. Auranen and Nieminen (Auranen & Nieminen, 2010) found that the cost of a publication, measured in terms of the USD in year 2000, is relatively constant over time, but varies from 70,000 to 170,000 across countries. This spread has to do with different tax rates across countries, but it is difficult to further normalize for 0% tax. Zhi and Meng (Zhi & Meng, 2016) analyzed the life science scientific output of institutions during the 11$^{th}$ Five Year Plan (2006-2010), and found that the number of papers published per institution is roughly proportional to the funding level it received. In this paper, we argue that the number of publication as a proxy for research investment is useful, even though it is not entirely uncontroversial.

Next, to measure the total commercial value (realized or potential), we look at the total number of patents awarded on graphene. The earliest study on the commercial potential of patents was the Patent Utilization Study, where of the 600 questionnaire responses analyzed, 30% were found to be in current use (Rossman & Sanders, 1957), and 34% of the grant in use reported an average profit of 600,000 USD (Sanders, Rossman, & Harris, 1958). More recently, Griliches (Griliches, 1990) found the average patent utilization rate to be 55% on average, but as high as 74% for small companies and inventors. Morgan et al. (Morgan, Kruytbosch, & Kannankutty, 2001) found that the commercialization rates of US patents was 48.9% for industries, and 33.5% in the education sector. These are consistent with what Svensson (Svensson, 2007) found when examining a small data set of Swedish patents, that the commercialization rates were between 52% to 74%, depending on whether the patents were held by individuals, small companies, or large companies. Webster and Jensen (Webster & Jensen, 2011) found using survey data that about 40% of patents advanced to the commercialization stage. Wu et al. (Wu, Welch, & Huang, 2015) found using survey data from the 2010 National Survey on Intellectual Property in Academic Science and Engineering that among various factors, university inventions are more likely to be licensed if the inventors themselves have positive attitudes towards the commercialization of their research, and if they collaborate with industries, if they engage in follow-up research, and if the university has a technology transfer office. The overall rate of licensing is reported to be 50.4%.

On the problem of patent value, Cutler (Cutler, 1987) analyzed 248 patents resulting from National Science Foundation Engineering Program grants, and found that 7 of them reported annual royalties between 10,000 USD to 250,000 USD. Cutler further estimated that the total long-term royalties from these 248 and related patents can be as high as 52.5 million USD, and the market value of their associated products may be ten or twenty times more. Subsequently, more groups employed direct survey of patent holders to ascertain the values of their patents, and developed regression models based on a variety of factors to estimate the values of other patents. Reitzig (Reitzig, 2004) classified these models into three generations: those based on citations (Carpenter, Cooper, & Narin, 1980; Trajtenberg, 1990; Harhoff, Scherer, & Vopel, 2003), those based on procedures (Putnam, 1997), those using the full-text information (Lanjouw & Schankerman, 1998; Tong & Frame, 1994). From this review of the literature, we see that the estimation of patent value is a very difficult problem. However, patent databases are reasonably complete. In contrast, it is even more difficult to compile a list of all graphene products, their prices, and their annual sales. Therefore, we settle for analysis of patents even though it is not the most direct measure of market value.

## 3. Data & Methods

### 3.1. Data

We use 'graphene' as a keyword to search the Web of Science ("Web of Science," n.d.) and downloaded the bibliographic records of 135,617 journal publications related to graphene. We realize that there might be many more graphene papers published in journals not indexed by the Web of Science, but believe that the most important papers should be included in this collection. For our analysis, we further exclude records without DOI number or publication year. We then refer to the remaining 127,546 records as the Graphene Science (G-S) collection.

Many papers in G-S collection also cite papers on nanotubes. Therefore, we also use this keyword to search and download the bibliographic records of 217,143 journal publications related to nanotubes from the Web of Science. Again, we believe the most important papers are included in this incomplete collection, which we refer to as the Nanotube Science (NT-S) collection after removing invalid results.

Over the last few years, scientists originally working on graphene have started exploring other two-dimensional (2D) materials. Some of these are based on transition metal dichalcogenides (TMD), while others have focused on transparent metal oxides (TMO) materials. Therefore, we also searched for and downloaded the 2D Materials Science (2DM-S), Transition Metal Dichalcogenides Science (TMD-S), Transparent Metal Oxides Science (TMO-S) collections. Details of these five collections (which represent the scientific inputs) are shown in Table 1.

From Derwent Innovation ("Derwent Innovation," n.d.), we searched for and downloaded matching collections of patent records: Graphene Technology (G-T), Nanotube Technology (NT-T), 2D Materials Technology (2DM-T), Transition Metal Dichalcogenides Technology (TMD-T), Transparent Metal Oxides Technology (TMO-T) collections (which represent the technological outputs), as shown in Table 1. In the patent collections, records without publication number or publication year are also excluded.

Table 1: Journal publication records from the Web of Science and patent records from Derwent Innovation found using the five keywords: "graphene", "nanotube", "2D material", "transparent metal oxide" and "transition metal dichalcogenide".

| | | Graphene | Nanotube | 2D Material | Transparent Metal Oxide | Transition Metal Dichalcogenide |
|---|---|---|---|---|---|---|
| Web of Science | Number of records | 127,546 | 196,489 | 24,763 | 29,488 | 1,448 |
| | Period | 1991 to 2017 | 1992 to 2017 | 1982 to 2017 | 1948 to 2017 | 1972 to 2017 |
| Derwent Innovation | Number of records | 176,193 | 373,527 | 1,611 | 122,977 | 1,100 |
| | Period | 1986 to 2017 | 1976 to 2017 | 1907 to 2017 | 1951 to 2017 | 1976 to 2017 |

The ten collections built using these keywords contain overlaps, i.e. a paper in G-S may also be a member of NT-S, or a patent in G-T may also appear in TMO-T. In Table 2, we show the number of overlapping papers across the three journal collections G-S, NT-S, (2D+TMO+TMD)-S, and in Table 3 the number of overlapping patents across the three patent collections G-T, NT-T, (2D+TMO+TMD)-T. In the rest of paper, when we need to classify a journal paper, journal reference, patent, or patent reference, we check it against the collections always in the order G-S(T), NT-S(T), and (2D+TMO+TMD)-S(T). This means that a journal paper that is in the G-S collection will be classified as belonging to this collection, even if it also can be found in the NT-S collection. Similarly, a patent that is first found in the NT-T collection will be classified as belonging to this collection, even if it also belongs to the (2D+TMO+TMD)-T collection. We realized that this is not an ideal treatment of the data, but there are no easy alternatives for such overlapping data sets.

Table 2: Overlapping records in the G-S, NT-S and (2D+TMO+TMD)-S collections.

| Number of overlapping records | G-S | NT-S | (2D+TMO+TMD)-S |
|---|---|---|---|
| G-S | - | 34,427 | 6,168 |
| NT-S | 34,427 | - | 2,235 |
| (2D+TMO+TMD)-S | 6,618 | 2,235 | - |

Table 3: Overlapping records in the G-T, NT-T and (2D+TMO+TMD)-T collections.

| Number of overlapping records | G-T | NT-T | (2D+TMO+TMD)-T |
|---|---|---|---|
| G-T | - | 77,835 | 10,754 |
| NT-T | 77,835 | - | 15,868 |
| (2D+TMO+TMD)-T | 10,754 | 15,868 | - |

### 3.2. Method

#### 3.2.1. Cleaning non-patent references of patents

Papers in the science collections are identified by their DOIs, whereas patents in the technological collections are identified by their publication numbers. Patents cite other patents, but also cite non-patents. Some of these non-patents are journal publications. To a lesser extent, journal papers also cite patents. Patents in the references of other patents can be easily identified by their publication numbers, while journal papers in the references of other papers can be easily identified by their DOIs. However, this is not true for journal paper references of patents, as well as patent references of journal papers. To highlight the severity of this problem, consider the number of G-T patents with publication code CN (68,816), which is approximately double that of G-T patents with publication code US (34,435). The numbers of non-patent citations of the CN and US sub-collections are 40,398 and 395,513 respectively. However, we recognize only 122 DOIs in the non-patent references of CN patents, but 142,770 DOIs in the non-patent references of US patents. We checked and found that this is the result of incomplete details (in particular the DOIs of journal papers) in the non-patent references. One of our first tasks is therefore to map these journal citations to their DOIs. We do this by sending each non-patent citation (less explicit references to DOIs) as a query request to CrossRef using their Metadata API, and record the most relevant result that CrossRef returns.

Naturally, we cannot simply take CrossRef's answers as the ultimate truth. In this second part, we verify our list of non-patent references against the list of most relevant results returned by CrossRef. We do this verification in four different ways. CrossRef returns the results in JSON (JavaScript Object Notation) format, allowing us to easily extract the authors and the titles. First, we compare the full non-patent reference against the combination of all authors and title by computing the Levenshtein edit distance (Tanaka, 2013/2019) between them. In some of these non-patent references, the list of authors may be missing or incomplete, or the title may be missing or incomplete. Therefore, we make a second comparison of the full non-patent reference against the combination of first author and title. In these two comparisons, the Levenshtein edit distances are large because the full reference contains more information than just the authors and title. Therefore, in the third comparison we truncate the non-patent reference after the last quotation mark (this should be at the end of the title) and compare this truncated result against the combination of all authors and title. Finally, we compare the truncated result against the first author and title.

However, the string lengths are different in the different combinations, so it is not fair to directly compare their Levenshtein edit distances. Instead, we compute the ratios of Levenshtein edit distances over the sum of lengths of the two strings being compared, and take the largest ratio to be our figure of merit. That is, if this ratio is small, we are confident that CrossRef has found the correct record. On the other hand, if this ratio is large, then CrossRef has found the wrong record. To determine the threshold for ratios we should accept, we created three test samples of 200 random references whose DOI numbers we know, as shown in Figure 2. For each reference, we delete everything from 'DOI' onwards, and post a query to CrossRef. We then record the smallest Levenshtein ratio between the CrossRef result and the truncated reference, as well as whether the CrossRef DOI is correct. The threshold Levenshtein ratio is determined using Otsu's algorithm (Otsu, 1979) to be 0.281, as shown in Figure 3.

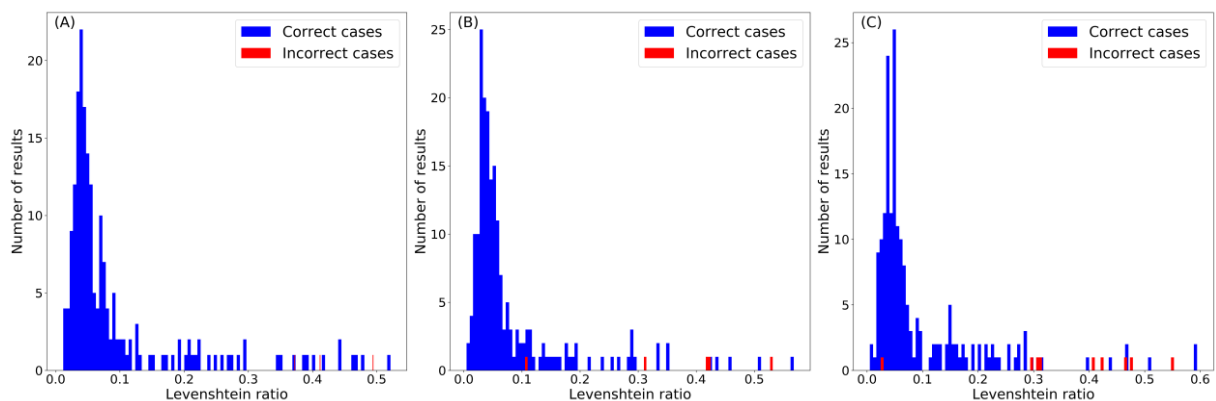

Figure 2: Histograms of the number of correct (blue) and incorrect (red) results from CrossRef for the three test samples, each having 200 random references.

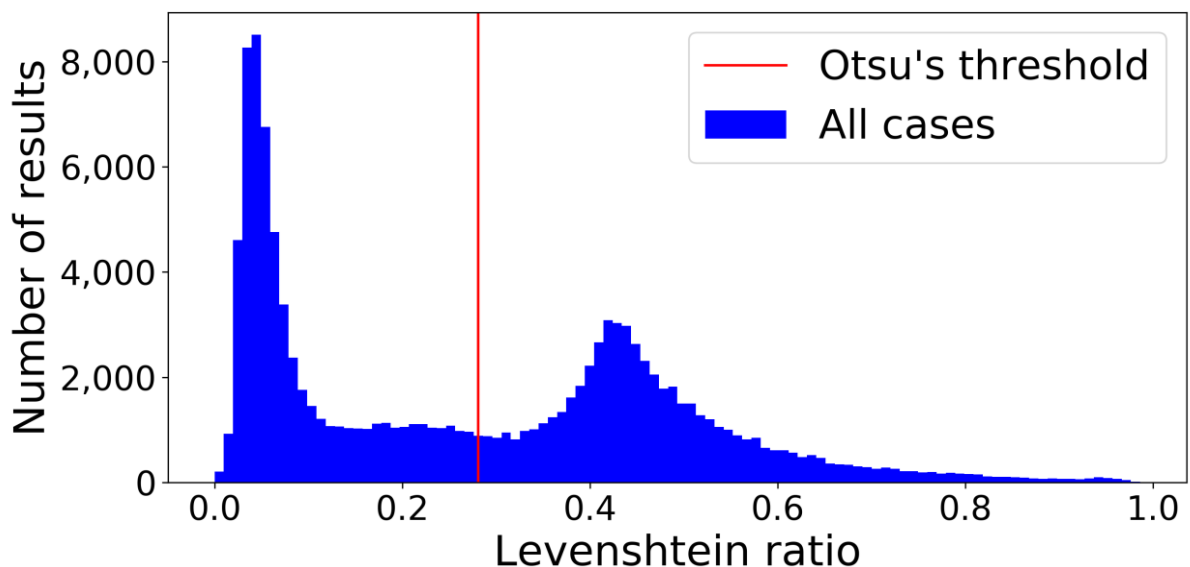

Figure 3: Histogram of results from CrossRef and the Otsu's threshold of accuracy. The bimodal histogram is expected to be the combination of two distributions, where the distribution with smaller peak corresponds to the correct cases, as observed in Figure 2. The Otsu's threshold separates the two classes with minimum variance.

3.2.2. Cleaning patent references of journal papers

Patent numbers in journal references may not include their regional code (first two characters, such as US, CN) or kind codes (last two characters, such as A1, B1). To determine whether such patent references are inside our collections, we ignore the kind codes, and check the lengths of their identifying numbers. If the length is less than 4, the reference is not a patent. For a longer identifying number, we check it against the patent numbers in the G-T, N-T, 2D-T, TMO-T and TMD-T collections, to look for a perfect match. Else, we regard this reference as an Others patent.

### 3.3. Aggregate Inflow-Outflow Analysis

After cleaning the data set, we are now ready to answer the question on how much ROI we are generating through funding investment into graphene research. First, we look at the aggregate level. For the 127,546 papers related to graphene published, we count the total number of papers cited, and decide whether these citations are also graphene papers, or nanotube papers, or 2D materials (including TMO and TMD) papers, or others (unrelated to the three class of materials). This represents the cumulative total scientific input to date to our scientific understanding of graphene. We also count the number of papers citing these 127,546 graphene papers, and again sort them into graphene papers, nanotube papers, 2D materials papers, or others. This represents the cumulative total scientific output to date from our scientific understanding of graphene. We then calculate the average number of papers cited by a graphene paper, the average number of citations received by a graphene paper, as well as the scientific value added (average number of citations received divided by average number of papers cited), taking into consideration which areas are contributing and which other areas are benefiting.

We then do the same for the 176,193 graphene patents, and count the total number of patents cited (sorted according to G-T, NT-T, (2D+TMO+TMD)-T and Others) and the total number of patents citing the graphene patents (sorted according to G-T, NT-T, (2D+TMO+TMD)-T and Others). These represent the cumulative total technological input to graphene technology to date and the cumulative total technological output from graphene technology to date respectively. By comparing the average number of cited patents to the average number of citing patents, we get a sense of whether graphene in its sum total as a technology is in a disruptive phase or a consolidating phase (Funk & Owen-Smith, 2017).

### 3.4. Temporal Analysis of Graphene Science and Technology

In calculating the simple scientific value adding, we have neglected the fact that the newer papers are still accruing citations. In fact, even old graphene papers can continue to be cited, and if the citing papers are related to graphene, these will add to the scientific input side of the equation. Therefore, we plot the number of graphene papers published in each year, the average number of references for graphene papers in each year, and the average number of citations of graphene papers in each year. Following this, we break down the graphene paper references and citations into components, depending on which collections they belong to. By building a model of citations based on the average citation profile of graphene papers, we can estimate the level of scientific interest on a year-to-year basis.

We then do the same temporal analysis for patents, by plotting the yearly number of graphene patents, the yearly average number of patent references, and the yearly average number of patent citations, as well as how their components change from year to year. We also estimate the level of technological interest in graphene, using the same method for estimating scientific interest.

### 3.5. Geographical Analysis of Graphene Science and Technology

Graphene, as an area of scientific inquiry, does not depend only on time, but also on space. In particular, we would like to look at the geographical features of graphene science and technology, to understand what is happening in the scientific contest between nations (Cimini, Gabrielli, & Sylos Labini, 2014).

3.5.1. Regional credit allocation

Unlike patents whose publication numbers include the regional codes of their registration offices, journal articles can be the results of the collaborations of many scientific researchers from different countries. In order to identify the top players in graphene science and technology, we need to assign regional contributions for the 127,546 G-S papers published from 1991 to 2017. We use the method proposed by Shen and Barabási (Shen & Barabási, 2014), which considers the co-authorship factor in allocating contributions among authors. The philosophy behind this algorithm is that it is not fair to split the credit evenly between all coauthors. Some authors contribute more and others less. Authors who contribute more are expected to write more papers on the same topics. Therefore we have to consider the collection of papers that are cited together frequently with the target paper. The proper credit that we should assign to an author of target paper must then take into account how frequently he/she was cited in the collection, as well as how highly cited individual papers are in this collection. Once this author credit allocation is done, we then proceed to calculate the contributions by different regions by going through all the papers and all the authors, and assign this credits to different regions based on the affiliations of authors. For authors with more than one affiliation, we divide his/her credit evenly between all his/her affiliations. To be fair, we should allocate more credit to an institution an author spends more time at. Unfortunately this information is not readily available.

3.5.2. Citations flow

We select the six regions with the largest numbers of G-T patents, namely China, US, South Korea, WO, Japan, and EP, and analyze citation flows between them. These regional flows can be calculated easily because the cited patents all contain their respective regional codes. We do not need to check whether the cited patents belong to any of our collections. Suppose the total number of patent references in the G-T patents is $N$, and the numbers of such patent references belonging to the six regions are $n_1, \ldots, n_6$. The size of the circle in Figure 11 representing region $i$ is proportional to $n_i$. A directed link from node $i$ to node $j$ means that patents in region $j$ cited patents in region $i$. The thickness of this link is proportional to the fraction of region $j$ patent references that are region $i$ patents.

For journal publications, we again select the top six regions (China, US, South Korea, India, Iran, and Japan) in terms of number of publications, using the affiliations of the first authors as the regions where the publications come from. Unlike for the case of patents, regional flows of journal citations are difficult to determine precisely from the references, because journal papers do not contain regional codes. More importantly, many papers nowadays are the results of international collaborations. For journal references outside of our collections, even if we know the first author, we do not know his or her affiliation(s). Therefore, we restrict ourselves to journal papers within our collections. The size of circle in Figure 13 representing region $i$ is the proportional to the number of journal papers that appear in the references of G-S papers whose first authors are from region $i$. If a first author has multiple affiliations, we split the reference count equally for each affiliation, whichever region it belongs to. A directed link from node $i$ to node $j$ means that journal papers from region $j$ cited journal papers from region $i$. The thickness of this link is proportional to the fraction of region $j$ journal references that are region $i$ papers.

## 4. Results

### 4.1. Aggregate Inflow-Outflow Analysis

As shown in Table 1, there are 127,546 papers related to 'graphene' published between 1991 and 2017, while there are 176,193 patents related to 'graphene' awarded between 1986 and 2017. The ratio is 176,193/127,546 = 1.38. This is an indicator for the commercial value (as measured by number of patents) of the science (as contained in the body of published papers). To convert this number to an actual estimate of the ROI, grant agencies must make a more precise estimate of the average cost $x$ of production of graphene papers, and track the commercialization of patents that come out from these studies, to see the average commercial potential $y$ of graphene patents. They should then multiply $y$ by 1.38 (because each journal paper 'generates' 1.38 patents), and then divide by $x$, to get the ROI. If this final ratio is less than one, then there is less return on the investment, and funding agency may want to stop funding. However, as we have noted in Section 2, the ultimate commercial value of a patent or a scientific paper is difficult to estimate. Fortunately, we have other collections that we can compare the graphene number against. For 'nanotube' (using data from 1992 to 2017), we find that the ratio is 1.90, whereas for 'transparent metal oxide' (using data from 1951 to 2017), the ratio is 4.17. This ratio appears to depend sensitively on the research field in question, and can potentially change with time, making a temporal analysis necessary.

But before doing the temporal analysis, we know that papers cite other papers, and patents cite other patents. We therefore proceed to do an inflow-outflow analysis, where we expect the inflow to depend on the maturity of the field and constitutes core 'graphene' papers that almost all other 'graphene' papers will cite. Conversely, we expect the outflow to have to do with impact generated by the field. For example, if in future only graphene papers cite graphene papers, then the impact of graphene research would be very narrow and limited to graphene only. In contrast, if future citations of graphene papers come from outside of the collection of graphene papers, for example in other materials, in engineering, in medicine, then we think of graphene science as having broad impact on many areas of science and technology. For patents, we need a similar interpretation of inflow and outflow. This was provided by Funk and Owen-Smith (Funk & Owen-Smith, 2017), who call a patent *consolidating* if all patents that cite it also cite patents cited by the target patent, and *destabilizing* if all patents that cite the patent do not cite any of the patents cited by the target patent. The natural question to ask then is whether graphene technology is currently in a consolidating or a disrupting phase.

In Figure 4, we show the bowtie diagram for 127,546 graphene journal papers, which cited 5,312,228 references, and were cited by 4,122,164 papers. At the aggregate level, each graphene paper cites 41.6 references, and is cited 32.3 times. The ratio $32.3/41.6 = 0.78$ is an indication of the scientific value of graphene as a research field. Typically, a paper in an emerging field will cite few references in the field, but will be cited many more times as the field grows. Therefore, the ratio of citations/references will be significantly greater than 1. When we average over the whole field, this ratio will surely drop, but if the field is productive, then many papers in the field will have large citations/references ratio. At this stage of its development, the scientific field produces many papers that can be considered quantum leaps in our knowledge. When this ratio drops below one, many of the papers will not be heavily cited, suggesting that the field is no longer productive, and the typical paper represents only an incremental increase in our scientific knowledge.

If we now sort the references into the different collections, we find that 44.6% of the references of graphene papers are G-S papers, 11.4% of these references are NT-S papers, 0.9% of these references are (2D+TMO+TMD)-S papers, and the remaining 43.0% are from outside our collections. On the other hand, if we sort the citing papers into the different collections, we find that 57.5% are G-S papers, 5.3% are NT-S papers, 1.2% are (2D+TMO+TMD)-S papers, and the remaining 36.0% are from outside our collections. If we look at NT-S papers cited by G-S papers and NT-S papers citing G-S papers, we have the ratio $218,475/605,594 = 0.36$. This tells us that as a science, graphene takes more from nanotube than it gives back. In contrast, if we look at (2D+TMO+TMD)-S papers cited by and citing G-S papers, we have the ratio $49,466/47,810 = 1.03$. Notwithstanding the fact that 2D materials as a field is still emerging, and the keywords are in flux, this suggests that graphene is giving more to 2D materials than it is taking. Finally, we look at Others papers cited by and citing G-S papers, to find the ratio $1,483,979/2,284,258 = 0.65$. Therefore, at the aggregate level, graphene does not contribute much to the growth of scientific research outside the few areas we have already identified.

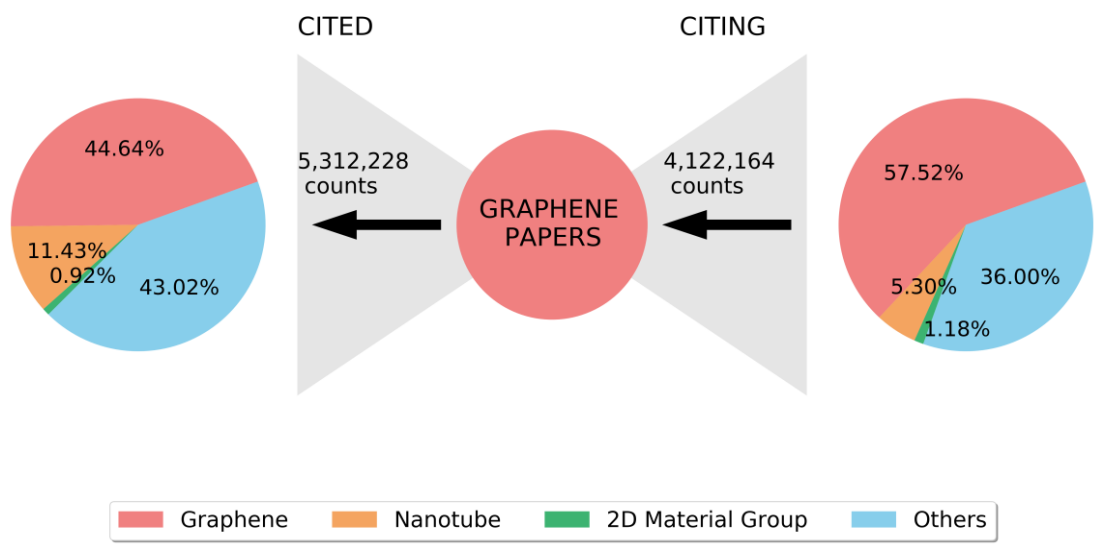

Figure 4: Bowtie diagram for journal paper citations of the G-S collection.

In Figure 5, we show the bowtie diagram of the 176,193 G-T patents, which have 1,458,760 patent references, and are cited by 416,550 patents. On average, each graphene patent cites 8.28 patents, and is cited by 2.36 patents. The output-input ratio is 0.29. This is an indication of the technological value of graphene, but as with G-S journal papers, it is difficult to understand the significance of 0.29 at the aggregate level. If we break the patent references down into collections, 15.8% of these are to G-T patents, 10.9% to NT-T patents, 1.8% to (2D+TMO+TMD)-T (which is reasonable, because level of interest in 2D materials became high later), the remaining 71.5% are to patents not in our collections. For the 416,550 citing patents, 41.5% are G-T patents, 14.1% are NT-T patents, 0.9% are (2D+TMO+TMD)-T patents, and the remaining 43.5% are outside our collections. This tells us that in contrast to graphene science, graphene technology draws inspiration from a broader literature than we had imagined, aggregating technologies developed elsewhere, and also has a broader impact on other technologies. In a previous study, Funk and Owen-Smith (Funk & Owen-Smith, 2017) examined 2.9 million US utility patents granted between 1977 and 2005, and see if they are destabilizing (producing new technologies) or consolidating (integrating existing technologies). It would therefore seem that with 416,550 patents citing the collection of 176,193 G-T patents, which cited 1,458,760 patents, graphene technology is on average more consolidating than it is innovating. Naturally, to make such a conclusion we would need to check the temporal behaviors of publications, references, and citations.

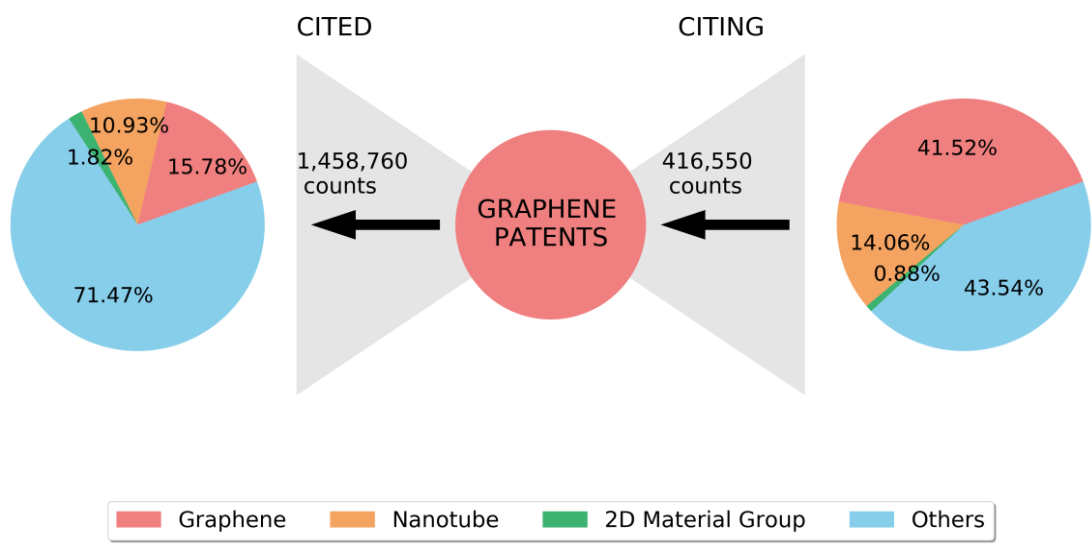

Figure 5: Bowtie diagram for patent citations of the G-T collection.

With the data we collected, it is also possible to examine G-T patents citing G-S papers, as well as G-S papers citing G-T patents. From the perspective of a grant agency, say from Singapore, who has funded the publications of $X$ G-S papers, and received in return $Y$ G-T patents, the investment is worthwhile if the ratio $Y/X$ is good. However, the actual answer may be subtler, as the Singapore G-T patents may not be citing the Singapore G-S papers. It might even be that the latter are cited by Japanese or US G-T patents, in which case it would be difficult to justify Singapore's investment of taxpayers' money. To answer this question, we look at Figure 6(A), to find 278,333 G-S papers cited by the G-T patents. Surprisingly, G-T patents cite more NT-S papers (15.7%) than G-S papers (14.8%). The most surprising result for us is how many papers outside of our science collections are being cited (68.9%). Now, why do graphene patents cite Others papers? Could these be papers that are cited by Others patents that the graphene patents also cite, and therefore the patent examiner insisted on including them? Or could these papers be probable technologies useful to the graphene patents, but are not yet patented?

Instead of answering our initial question, we find ourselves asking even more difficult questions. If a G-T patent cites a patent and also cites a journal paper that is cited by the latter, the patent and journal paper can be said to be correlated, and the G-T patent's citation of the journal paper is redundant. Conversely, if a G-T patent cites a journal paper that is not cited by its patent references, we say that the journal paper is anti-correlated with the patent references, and represent complementary ideas/technologies not found in the patent references. Since it is not possible to check the correlation between Others papers and Others patents cited by graphene patents, we test this on the NT-S, NT-T collections instead. To do this, 1,000 random G-T patents are chosen, in which we find 239 of them citing NT-T patents. In these 239 G-T patents, we find 32 of them having NT-S journal references, out of 77 that contain journal references. We then check for overlap between the NT-S journal references of G-T patents and NT-S journal references of their NT-T patent references. In 19 out of 32 G-T patents with NT-S journal references, there is no overlap between NT-S references. In the remaining 13 G-T patents, the average probability of finding a NT-S journal paper that is cited by a G-T patent and its NT-T patent references is $0.6 \pm 0.4$. Therefore, we find two groups in these 32 G-T patents, a consolidating group (13 G-T patents) that cites journal references of its patent references with high probability, and a disrupting group (19 G-T patents) that cites journal references not cited by its patent references.

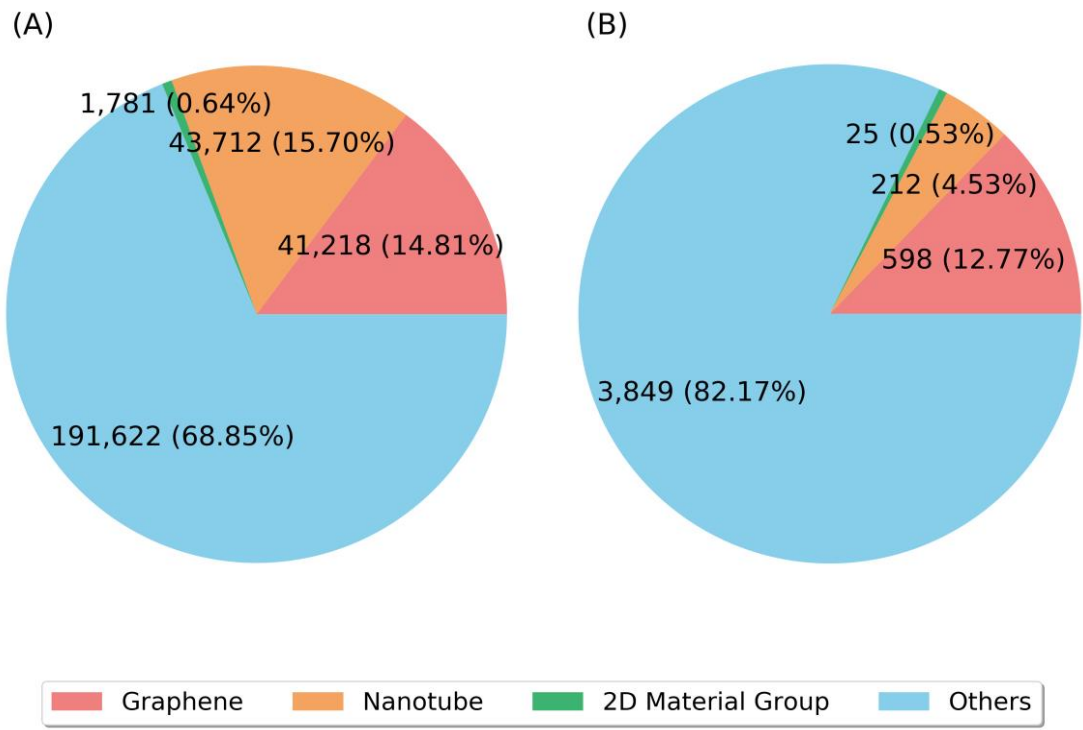

Figure 6: Components diagram from Graphene, Nanotube, 2D Material Group – combination of 3 collections: 2D material, Transparent Metal Oxide, Transition Metal Dichalcogenide and Others for (A) total papers cited by graphene patents and (B) total patents cited by graphene papers.

As expected, only 598 G-T patents are cited by G-S papers, as opposed to 278,333 G-S papers being cited by G-T patents. We also find 212 NT-T patents cited by the G-S papers, which is not surprising, because this number is less than that for G-T patents. What is more striking are the 3,849 patents outside of our collections being cited by G-S papers. Papers citing patents represent a reverse information flow from what we expect (science to technology), so this suggests a reverse information flow into G-S papers, from research area outside of those we have identified. A few examples of these Others patents cited by G-S papers include:

- US patent number 3330697 (M.P. Pechini (1967), "Method of preparing lead and alkaline earth titanates and niobates and coating method using the same to form a capacitor") is cited by G-S journal paper with DOI number 10.1007/s00339-004-2884-7 (González, M., Landa-Cánovas, A. & Hernández, "Pyrolytic and graphitic carbon: pressure induced phases segregated in polycrystalline corundum", M. Appl. Phys. A (2005) 81: 865);
- US patent number 7409759 (Sewell, H. (2008), "Method for making a computer hard drive platen using a nano-plate") is cited by G-S journal paper with DOI number 10.12989/SEM.2017.61.1.065 (Kiani, Keivan, Gharebaghi, Saeed Asil, & Mehri, Bahman. (2017). In-plane and out-of-plane waves in nanoplates immersed in bidirectional magnetic fields. Structural Engineering and Mechanics, 61(1), 65–76);
- Japan Patent Kokai, patent number 2008-169255 (Kuramoto N (2008) "Highly conductive polyaniline having excellent solubility and method for producing the same") is cited by G-S journal paper with DOI number 10.1007/s00396-015-3539-2 (Kuwahara, R.Y., Oi, T., Hashimoto, K. et al., "Easy preparation of graphene-based conducting polymer composite via organogel", Colloid Polym Sci (2015) 293: 1635).

## 4.2. Temporal Analysis of Graphene Science and Technology

At the aggregate level, the picture is worrisome, because whether we are looking at journal publications or patents, the number of citations is smaller than the number of references. If the interest level remains constant over time, we would expect the number of citations to grow proportionally as the number of publications (and hence the number of references). To put it simply, if each graphene publication (journal or patent) cites $n$ other graphene publications, and the number of graphene publications is $N$, the number of graphene references would be $nN$ while the number of graphene citations would be $nN$. This means that the ratio of citations to references would be 1. Naturally, an aggregate-level analysis is too coarse to check whether interest level is increasing or decreasing with time. The ideal way to check this is to look at publications, references, citations, and other indicators as functions of time.

In Figure 7(A), we show the number of graphene journal publications and patents over the years 2004 to 2017. From 2004 to 2010, the yearly number of papers is very close to the yearly number of patents. From 2011 onwards, the yearly number of patents started increasing faster than the yearly number of papers. This is opposite to what is found by Zou et al., (Zou et al., 2018) using Chemical Abstracts Service as their source. For the early years (2004, 2005) there are more graphene patents than graphene papers. We suspect this may be the result of aggressive tagging by Derwent. Now we have a time-resolved picture of the input.

For the time-resolved picture of the output, we could look at the numbers of papers citing G-S papers of various years, and also the numbers of patents citing the G-T patents of various. However, the early G-S papers and G-T patents have more time to accumulate citations compared to the latter G-S papers and G-T patents. Furthermore, the numbers of G-S papers and G-T patents published increase every year. Therefore, it is more meaningful to plot the average number of citations G-S papers (G-T patents) published in the various years, to answer the question whether the body of graphene science generating more citations over the years, or is it adding new papers faster? In Figure 7(B) and Figure 7(C), we show the average number of journal citations (including those outside of our collections) per paper for each year in the G-S collection, and the average number of patent citations (including those outside of our collections) per patent for each year in the G-T collection. We find a monotonic decrease over time of the average number of journal citations per paper. In contrast, the average number of patent citations per patent started out disparately smaller than the journal counterpart in 2004, but become nearly equal in 2017. In general, every G-S paper is generating fewer citations as time goes on. There are two possible reasons for this: (1) a paper published later has less time to accumulate citations, (2) a paper published later is less important on average to the future of the field. In contrast, the average citation of a G-T patent is roughly constant until about 2011/2012, when it starts to decrease. In the Discussion section, we simulate G-S journal publications and citations, as well as G-T patents publications and citations up to 2021 using the extrapolated yearly publications curve and the citation profile (the proportions of references of a typical paper, published $t$ years before the paper itself), and show that this decline is likely to be real, and not an artifact of the forward looking nature of the indicator. In fact, by calculating the ratio of the empirical average number of citations per paper and the simulated average number of citations, we argue that scientific interest in graphene increased from 2004 to 2010, and decreased thereafter.

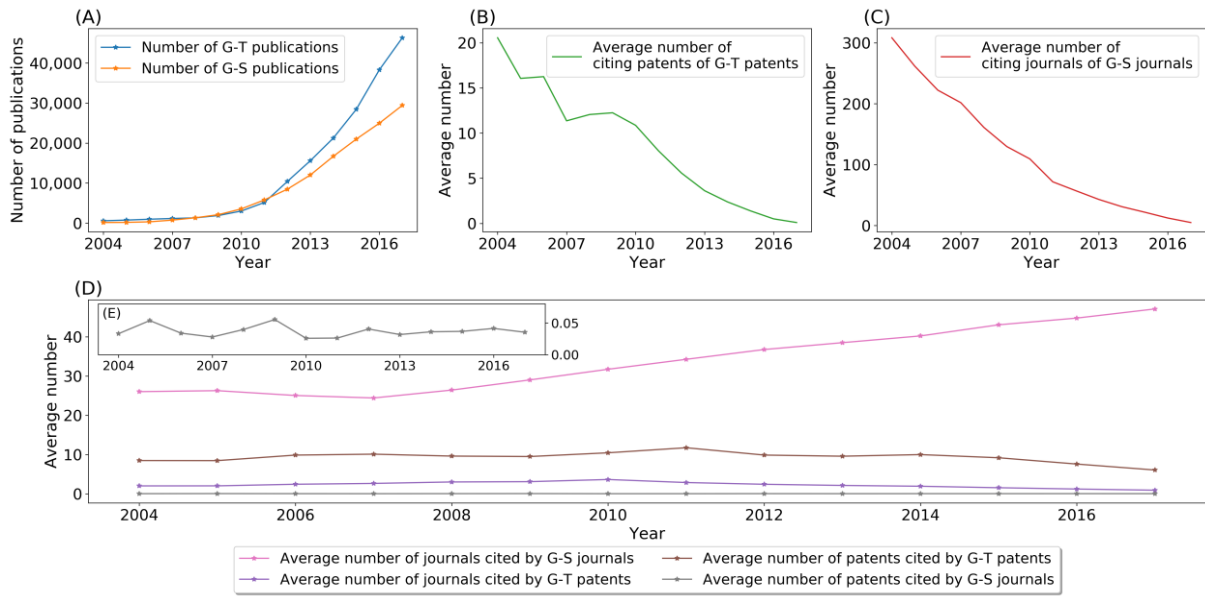

Figure 7: Distribution over time (2004 to 2017) in (A) Number of Graphene journals and patents publications; (B) Average journal citing counts of G-S collection; (C) Average patent citing counts of G-T collection; (D) and (E) Average journal and patent references of G-S and G-T records.

In Figure 7(D), we see the average number of journal articles cited by G-S papers increasing over the years. These include not only G-S papers, but also NT-S papers, and others. In contrast, the average number of patents (including patents from G-T, NT-T, 2D-T, TMO-T, TMD-T and others) cited by G-T patents is roughly constant over time. We also look at the average number of journal articles cited by G-T patents, and the average number of patents cited by G-S papers (shown magnified in Figure 7(E)), and find that these are also roughly constant over time.

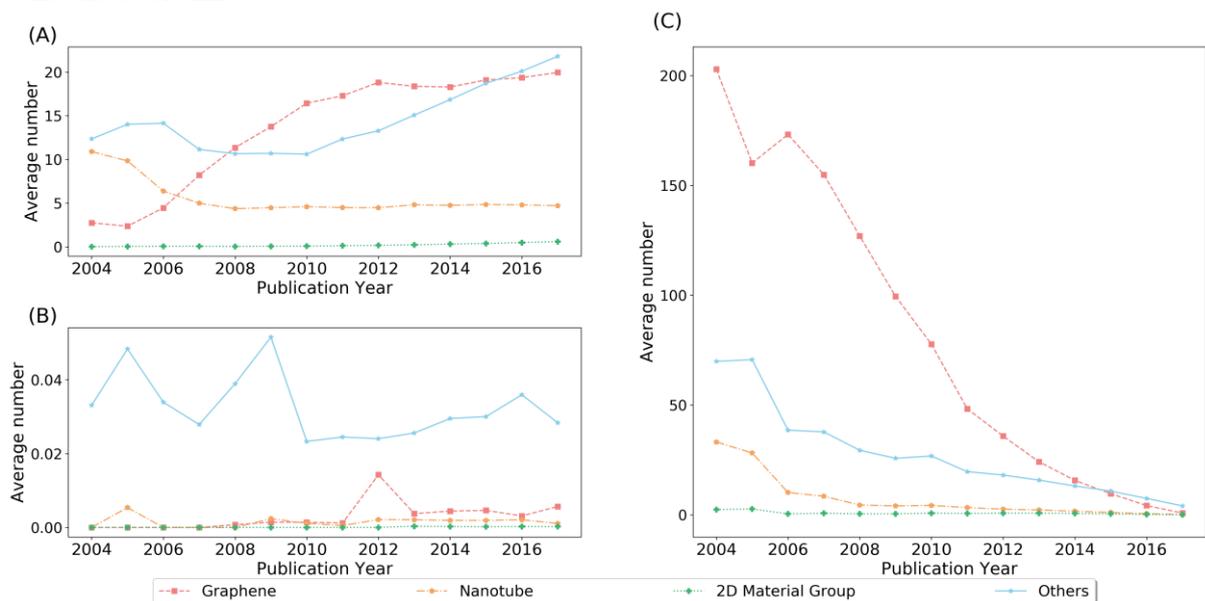

Figure 8: Citations components distributions of G-S collection from Graphene, Nanotube, 2D Material Group and Others categories in average number for (A) cited papers, (B) cited patents and (C) citing papers.

In Figure 8(A), we break up the references of G-S papers into the collections. Those references that are part of the G-S, NT-S, (2D+TMO+TMD)-S collections are shown in red, orange, and green respectively. Other references that are outside these collections are shown in blue. In the early years, there are more NT-S references than G-S references in G-S papers, but in G-S papers published after 2007 there are more G-S references than NT-S references. We do not know what the Others references are, but the number of G-S references started lower than Others references in 2004, overtakes the Others references in 2008, and went below the Others references in 2015. This suggests that there is a 'Golden Era' for graphene research between 2008 and 2015. There are no surges in this graph for 2D Materials, which is supposedly the next hot thing. Intriguingly, we see in Figure 8(B) that most of the patent references of G-S papers are outside of our collections.

From Figure 8(C), we see that most of the journal papers citing papers in G-S collection are themselves from the G-S collection. Surprisingly, the next most common category of journal papers citing G-S papers are from outside of our collections, instead of journal papers from the NT-S collection (which comes in third), or journal papers from the (2D+TMO+TMD)-S collection (which come in last). For each collection, the average number of papers citing the G-S collection is decreasing rapidly over time. Even if we explain this in terms of the rapid growth in the number of G-S papers over the years, a typical G-S paper will still be cited less.

In Figure 9(A), we break up the patent references of our G-T patents into the various collections. Most of these patent references are outside of our collections (Others), over all the years we have data on. In fact, up till 2012, the second most common patent references of the G-T patents are from the NT-T collection, before G-T patents take over as the second most common patent references of G-T patents. Figure 9(B) shows the distribution of papers cited by G-T patents. Again, the dominant component is Others and the second most dominant component is NT-S. The NT-S component peaks in 2008, while the Others component peaks in 2010. G-S papers overtake NT-S papers as the second most prominent component in 2013.

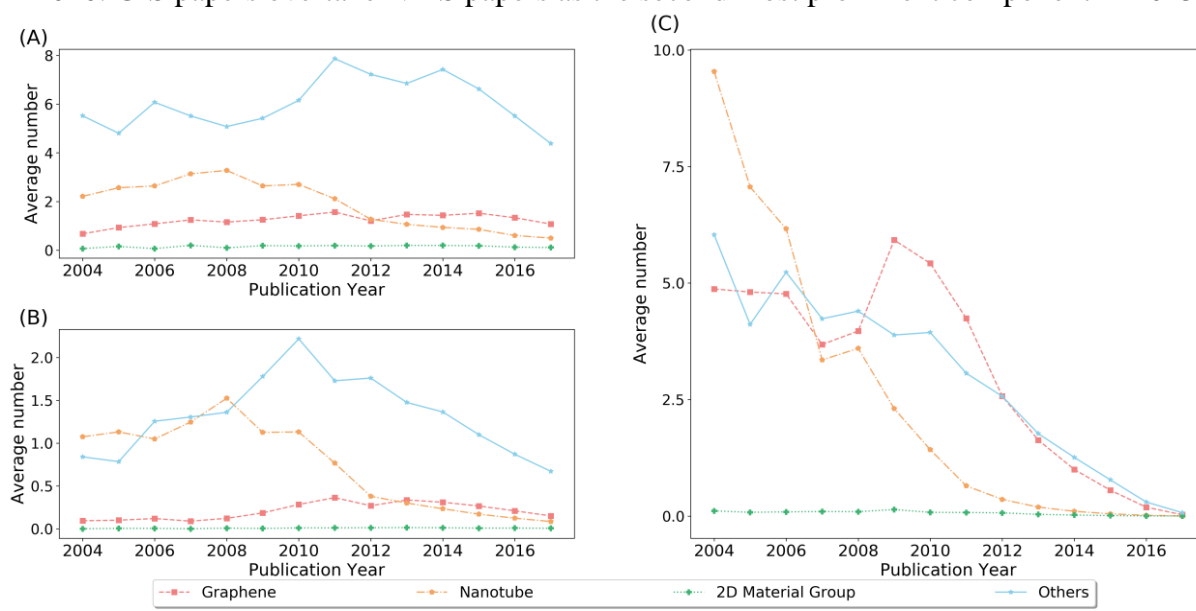

Figure 9: Citations components distributions of G-T collection from Graphene, Nanotube, 2D Material Group and Others categories in average number for (A) cited patents, (B) cited papers and (C) citing patents.

In Figure 9(C) we see that in 2004, most of the patent citations to G-T patents come from NT-T patents, follow by Others patents. The shares of NT-T and Others patents decrease with time, and in 2008 the average number of G-T patents citing G-T patents become greater than those from NT-T and Others. Unlike clear signs from the journal references of a 'Golden Era' in graphene science from 2008 to 2015, the signs of a similar 'Golden Era' in graphene technology can only be seen from the patent citations. The start of this technological 'Golden Era' is also 2008, but because citations can change over time we cannot be sure it has ended based on data up to 2017.

### 4.3. Geographical Distribution (Patents vs Papers)

Finally, we look at how graphene science and technology is generated in different geographical regions of the world. In Figure 10, we show the cartogram for total journal credits in the G-S collection, using the algorithm by Gastner and Newman (Gastner & Newman, 2004). We see that China is the largest source of graphene papers over the period 1991 to 2017. Other important regions include the US, Taiwan, Japan, Korea, Europe and India. In this figure, the contributions from different regions are shown clearly, and they are not equal.

In comparison, for graphene technology, we see from Figure 11 that there are four major players: China, US, Korea, Japan. In graphene science, Europe is also an important player, but in this figure it is not enlarged because we cannot assign patent credits to individual European countries.

Next, we examine the geographical patterns in patent-patent citations. In Figure 12, we see that US, China, Korea, Japan patents cite mostly patents from their own regions. On the other hand, WPO and EPO patents cite mostly US patents. Beyond regional self-citations, we find Korean patents citing more Japan patents than US and China patents. In Table 4, we show that this insular citation pattern is very likely due to the languages used in different patent

offices

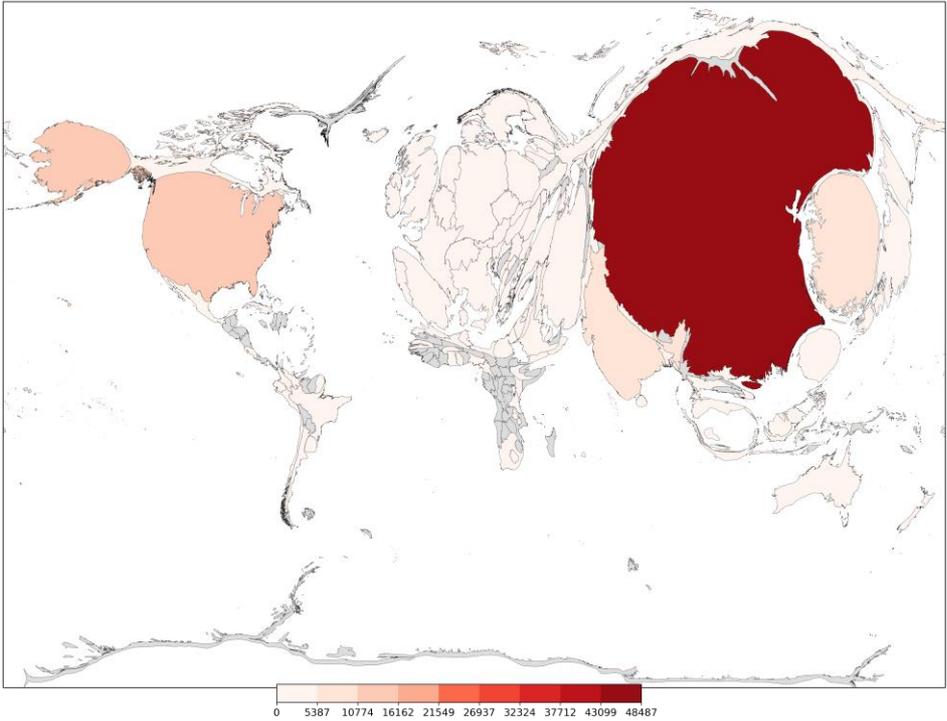

Figure 10: Cartogram for total journal credits in the G-S collection, showing the most important contributing regions enlarged relative to their natural sizes.

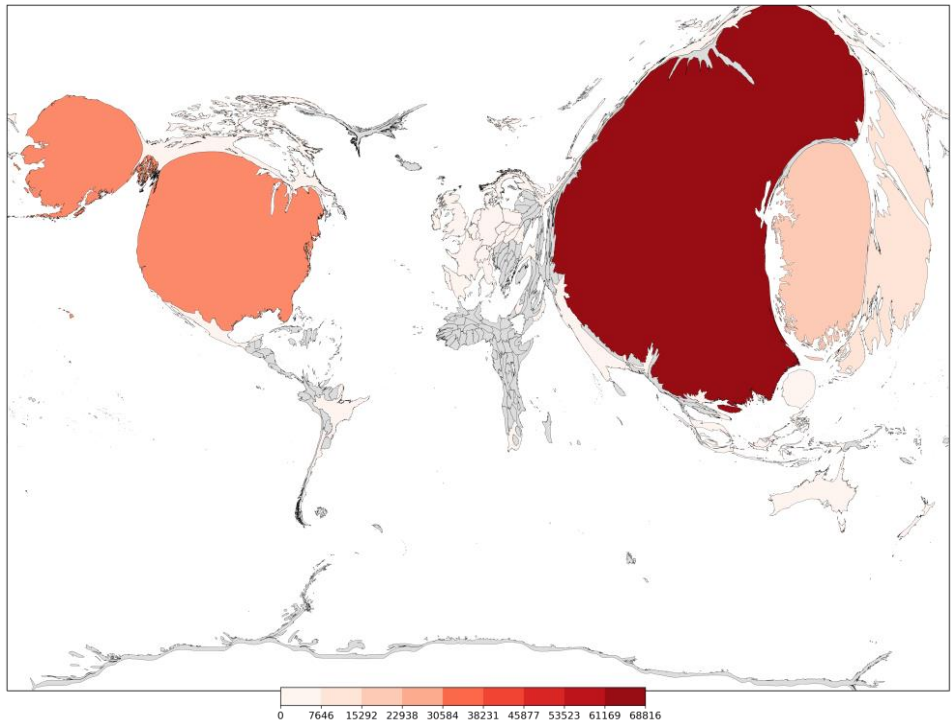

Figure 11: Cartogram of the total number of published patents in the G-T collection, showing the most important contributing regions enlarged relative to their natural sizes.

Table 4: The languages of patents filed in the various regional patent offices as well as the World Intellectual Property Organization (WO).

|    | ZH     | EN     | KO     | JA     | DE  | FR  | ES | PT | RU | XX |
|----|--------|--------|--------|--------|-----|-----|----|----|----|----|
| CN | 68,815 | 1      | -      | -      | -   | -   | -  | -  | -  | -  |
| US | -      | 34,435 | -      | -      | -   | -   | -  | -  | -  | -  |
| KR | -      | -      | 19,919 | -      | -   | -   | -  | -  | -  | -  |
| WO | 926    | 10,523 | 1,438  | 2,339  | 283 | 401 | 56 | 12 | 9  | 1  |
| JP | -      | 3      | -      | 15,161 | -   | -   | -  | -  | -  | -  |
| EP | -      | 8,037  | -      | -      | 358 | 390 | -  | -  | -  | -  |

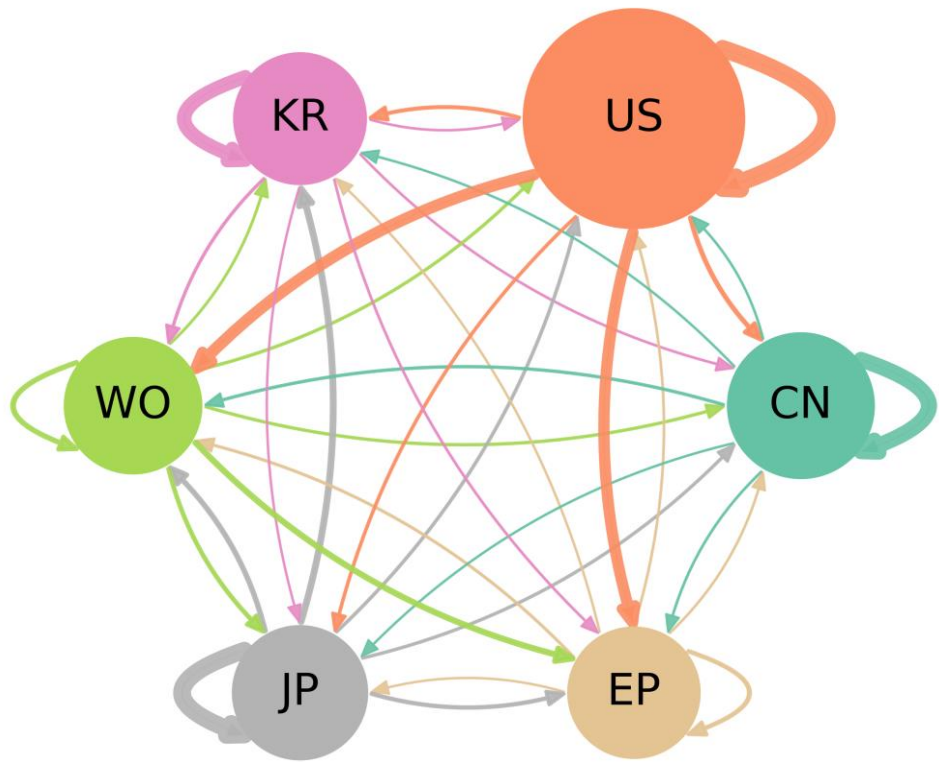

Figure 12: Patent citation flow of top six regional indices in total number of published G-T patents.

The picture shown in Figure 13 for journal-journal citations is different. Instead of US patents attracting the most patent citations, we now have Chinese journal papers attracting the most journal citations. Moreover, the journal citation patterns have a decidedly more global character. Instead of the regions predominantly citing themselves for patents, we see Chinese journal papers are cited strongly by South Korean, Indian, and Iranian journal papers, and also less strongly by US and Japanese journal papers. Similarly, US journal papers are cited strongly by South Korean, Indian, and Japanese journal papers, and also less strongly by Chinese and Iranian journal papers. Papers from other four regions are not strongly cited.

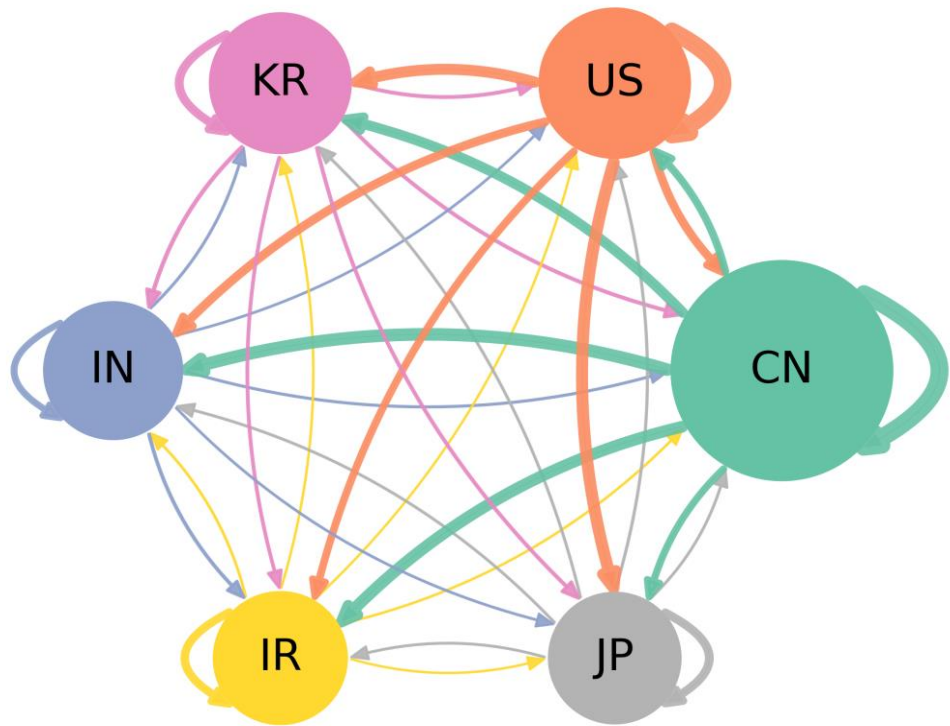

Figure 13: Paper citation flow of top six regional indices in total G-S papers credits from the first author's contribution.

We also examine the geographical distributions of journal paper references of G-T patents, for journal papers within our collections in Figure 14. For consistency, we consider patents from US, China, South Korea, and Japan because there are no journal papers that we can attribute to EP and WO. The patents from all four regions are more likely to cite journal papers from themselves, but Chinese and US journal papers are also highly cited by South Korean and Japanese patents. We do not examine patent references of G-S journal papers because there are simply too few of these.

Finally, we look at the competition between regions for graphene science and technology (Cimini et al., 2014). In Figure 15 we show the rankings of top ten regions in terms of journal publications from 1991 to 2017. The Web of Science assigns the keyword 'graphene' to journal publications as early as 1991, even though graphene became a hot topic only after 2004. For the pre-2004 journal publications, France appears to be the pioneer and led for 1991 and 1992, before US became the scientific leader. For the rest of period up to 2004, the leader of graphene science was essentially a contest between US and Japan. France remained in the top six until 2003, then fell out of top ten in 2004, and came back briefly within the top ten from 2005 to 2010, and from 2011 onwards it became permanently outside of top ten. The other region that saw its fortune ebbed is Japan. It is one of the earliest regions to pay attention graphene science, after a false start became one of the top two regions in graphene science between 1999 and 2008. After 2008 the scientific output of Japan in the area of graphene showed a steady decline. Graphene research in China started later, in 1996, had a period of ups and downs, but has seen a remarkable rise starting in 2006. It overtook US as the leader in Graphene science in 2011. South Korea and India, both of which dabbled in graphene science as early as 1997 and 1994 respectively, started their ascendances in 2008 and 2009, to become No. 4 and No. 3.

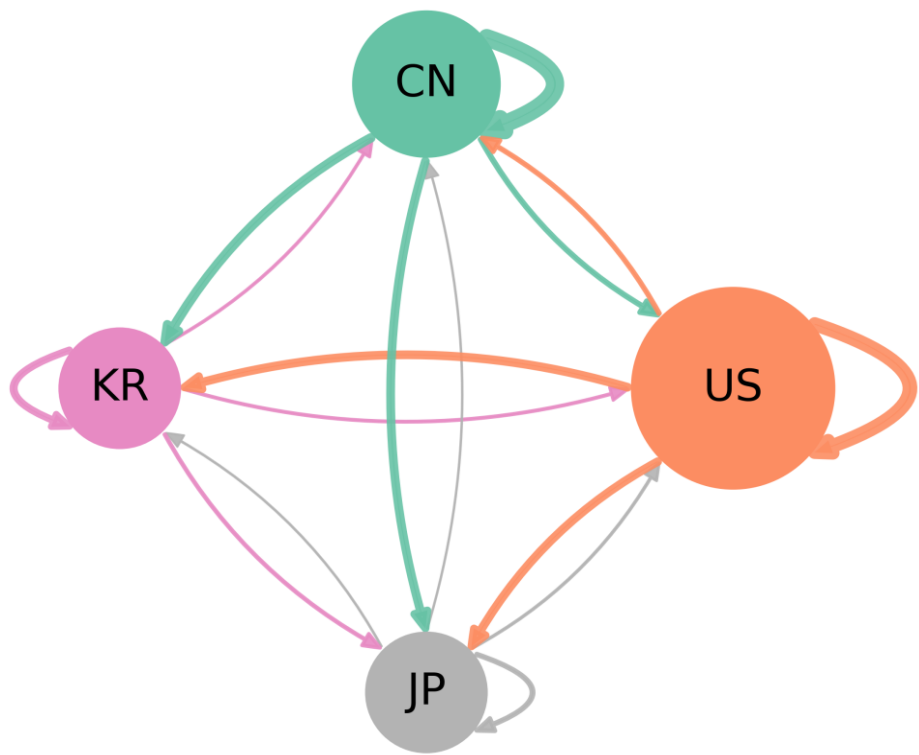

Figure 14: Paper citation flow of top 6 regional indices in total number of published G-T patents.

In contrast, the top ten regions for graphene technology are substantially different. According to Derwent's classification, the first graphene patent was published in 1986 in China. China was the leader of graphene technology up until year 2000, when it was displaced by Japan and US. China regained this leadership position in 2012, leading the runner up US by a big margin. Japan's position stared declining steadily from 2007. Contrast this to Japan's position in graphene science, which started to declining to 2004. The only clear rise is South Korea in 2008.

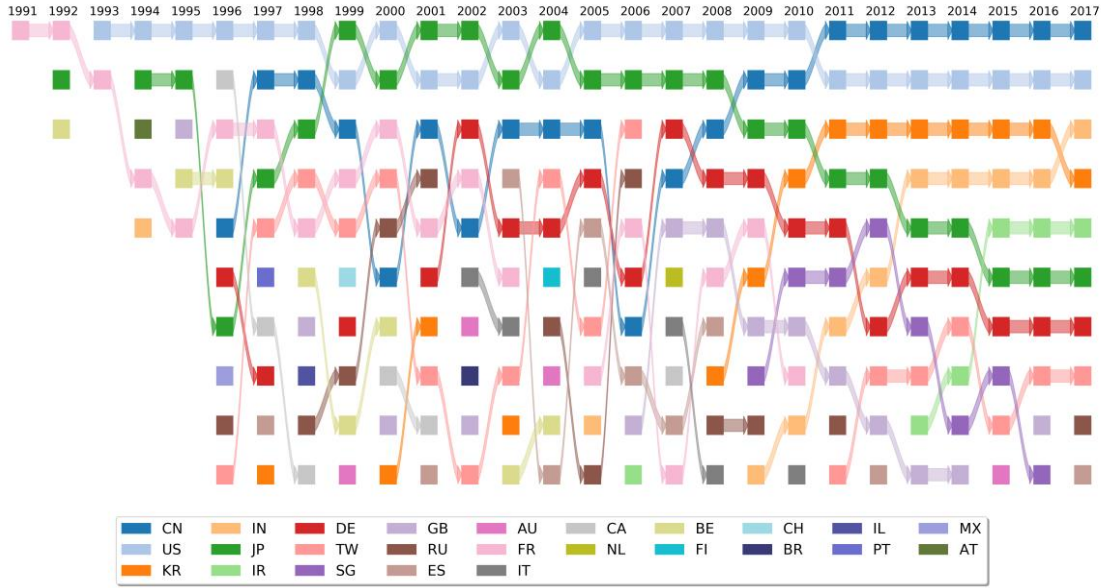

Figure 15: Top 10 regions and countries in journal credits every year in the G-S collection.

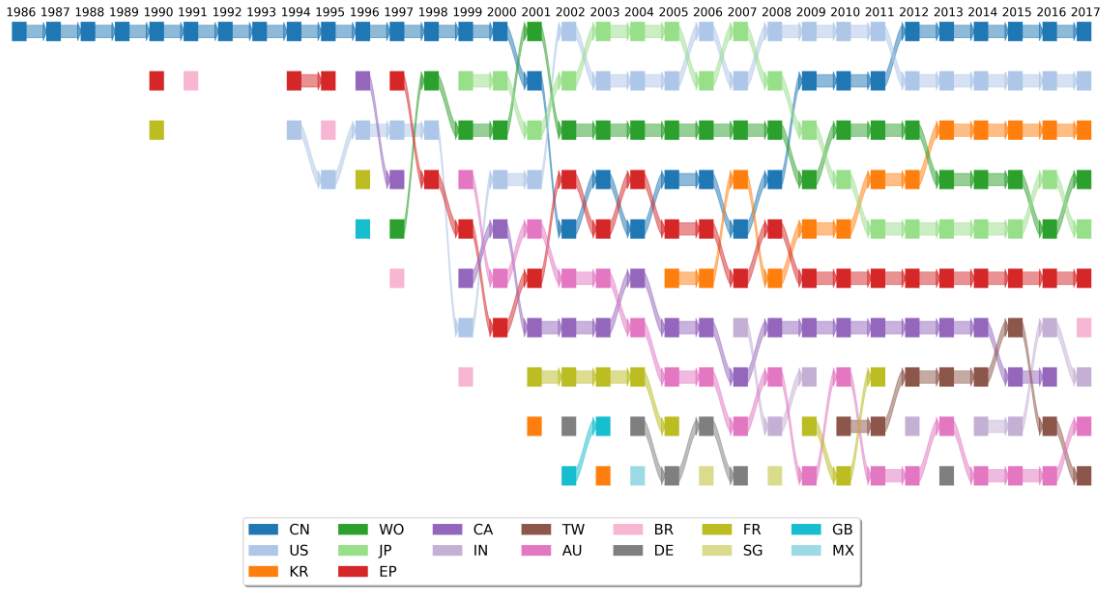

Figure 16: Top 10 regions and countries in number of patents publications every year in G-T collection.

## 5. Discussion

In Figures 7, 8 and 9, we plotted yearly quantities (for example, the yearly numbers of publications) that would no longer change with time, as well as forward-looking quantities like the average number of citations for journal papers and patents published in a given year. These average numbers of citations are calculated using the total number of citations accumulated up till 2017 (downloaded in 2018). The total number of citations will continue to increase with each passing year. Therefore, we have to be extra careful interpreting features seen in the average citation curves. The most important feature we care about in these curves is the decreasing trend. In other words, can the decrease we observe be explained by papers published in later years having less time to accumulate citations? Or is this decrease a result of scientists losing interest in the research field?

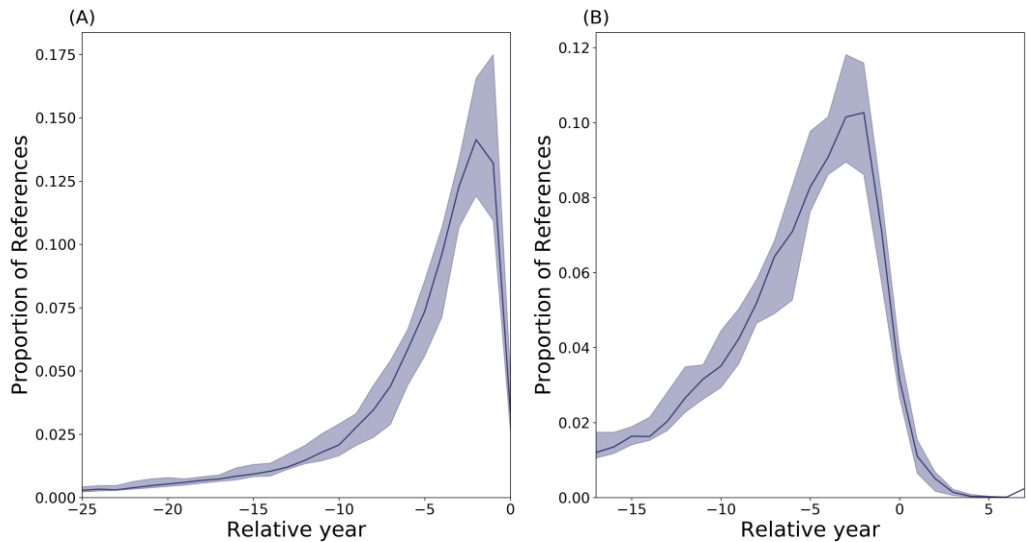

Figure 17: Citation profiles of (A) G-S journal papers and (B) G-T patents, showing the proportions of references from different years prior to their publications. We show the median curve as a solid line, while the shaded region is bounded by the $25^{th}$ and $75^{th}$ percentile curves. We limit the number of years prior to publications to 25 years, so that proportions add up to 90% of all references. We noticed there exists patents references that are published after the patent citing them, but we do not know all the reasons why these exist.

To discriminate between these two possibilities, we performed a simple simulation, based on the citation profiles of G-S papers and G-T patents shown in Figure 17. These citation profiles are relatively stable over time, and also do not vary much from research fields to research fields (Liu, Nanetti, & Cheong, 2017). First, we predicted how many journal and patent publications there will be from 2018 to 2021, by extrapolating the publication numbers from 2004 to 2017. We do this by assuming that the publication numbers for G-S and G-T are still in the exponential growth phase, and perform a linear regression (Tanagra, 2017) of the logarithms of publication numbers against the years, to obtain red dashed curves shown in Figures 18(A),(C). From the linear regression, we also obtained the 95% confidence intervals shown as red shaded regions shown in Figures 18(A),(C). Second, we assume each of these 'future' publications cite the same average number of references as publications between 2004 and 2017. We also assume that the proportions of 'future' references that are graphene publications are the same as those shown in Figures 4 and 5. Under these assumptions we generated a pool of aggregated 'future' graphene references for 2018, by using the predicted number of graphene publications in 2018, or the lowest and highest numbers of graphene publications according to the confidence interval. We then assign them to publication years 2017, 2016, … 2004 according to the 25%, 50%, or 75% citation profiles shown in Figure 17. In this way, we obtain a distribution for the 2018 contribution to our 'future' citations. We do the same thing for the pools of aggregated 'future' graphene references generated in 2019, 2020, 2021 and combine them to form the shaded region shown in Figures 18(B),(D). This is where the average citation curves could lie at the end of 2021.

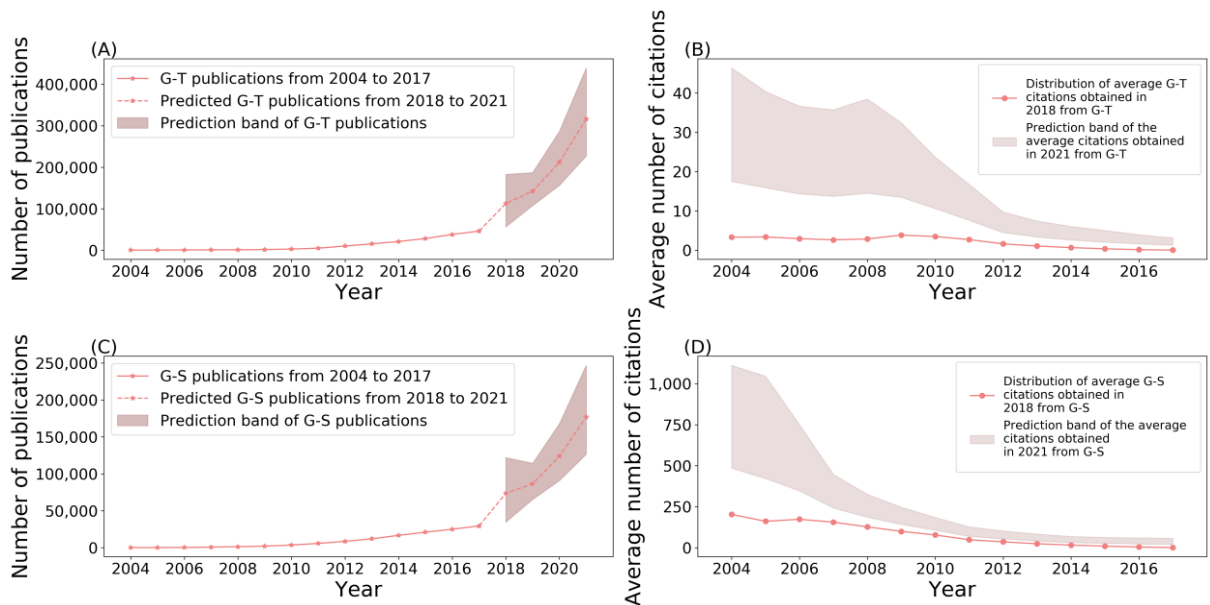

Figure 18: (A) The number of G-T patents published from 2004 to 2017 (solid curve). Yang et al. observe that the G-T patents was in a growth phase from 2011 to 2017 (Yang, Yu, & Liu, 2018) so we predicted the numbers (dashed curve) and uncertainties (shaded region) of G-T patents from 2018 to 2021 using the exponential growth formula (Ernst, 1997). (B) The average number of citations per G-T patent from within the G-T collection from 2004 to 2017 computed using patent records up till 2017 (solid curve), and the shaded region represents the possible average number of citations per G-T patent with additional simulated contributions between 2018 and 2021. (C) The number of G-S journal papers published from 2004 to 2017 (solid curve). We use the same exponential extrapolation method to predict the numbers (dashed curve) and uncertainties (shaded region) of G-S journal papers from 2018 to 2021. (D) The average number of citations per G-S journal papers from within the G-S collection from 2004 to 2017 computed using bibliographic records up till 2017 (solid curve), and the shaded region represents the possible average number of citations per G-S journal papers with additional simulated contributions between 2018 and 2021.

In Figure 18(B), we see that with citations up till 2017, the average number of citations per G-T patent is rough constant between 2004 and 2008, peaked at 2009, and thereafter decreases. With the simulated citations up to 2021, we see that while the average number of citations of G-T patent is still roughly constant, the peak has shifted to 2008. Even though we cannot say much about more recent years, the decreasing trend from 2008 to 2012 in the average number of citations per G-T patent is likely to be real. This is another indication that golden age of graphene technology is over, and interest in it is declining. In Figure 18(D), we see that the average number of citations per G-S journal paper start declining from 2004 onwards. For G-S journal papers, the signature of golden era is seen from the references instead of from the citations.

With the collections we have, we can also explore the cross-sectional characteristics of these citations. In Figure 19, we show the proportions of references from the G-S, NT-S, (2D+TMO+TMD)-S collections that are themselves from these collections. As expected, close to 50% of the references of G-S journal papers are from G-S collection. Surprisingly, references of NT-S journal papers that are from the NT-S collection make up only 34% of all references, while only 6% of references of (2D+TMO+TMD)-S journal papers are from this same collection. For the latter, we understand that 2D materials is a growing field of research with a relatively small literature base to cite from. For the former, we suspect the low intra-collection referencing is due to dwindling interest in nanotube research.

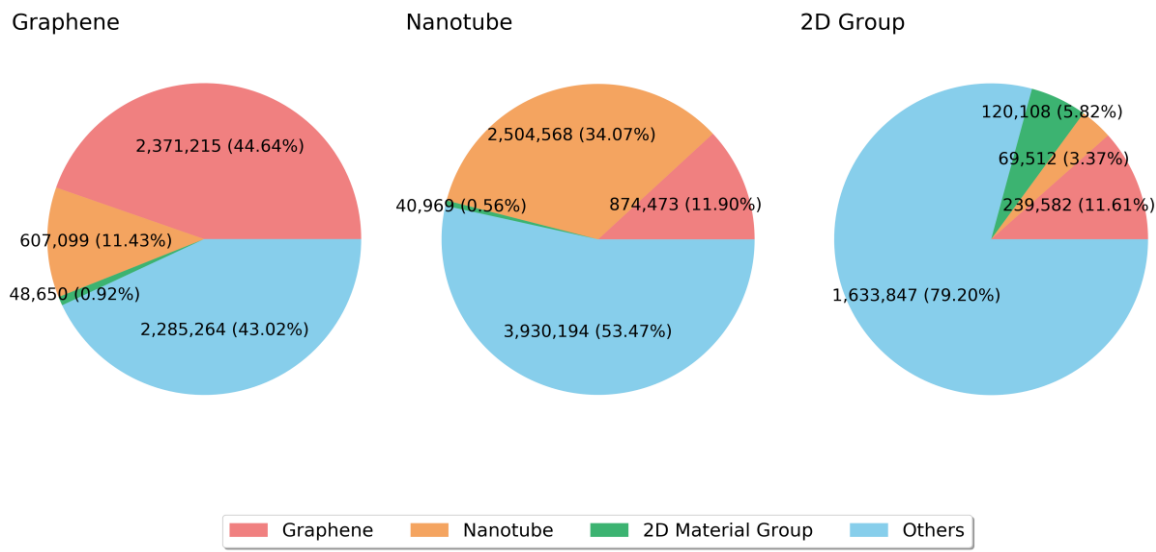

Figure 19: References of (left) G-S, (middle) NT-S, (right) (2D+TMO+TMD)-S journal publications sorted according to which collections they come from.

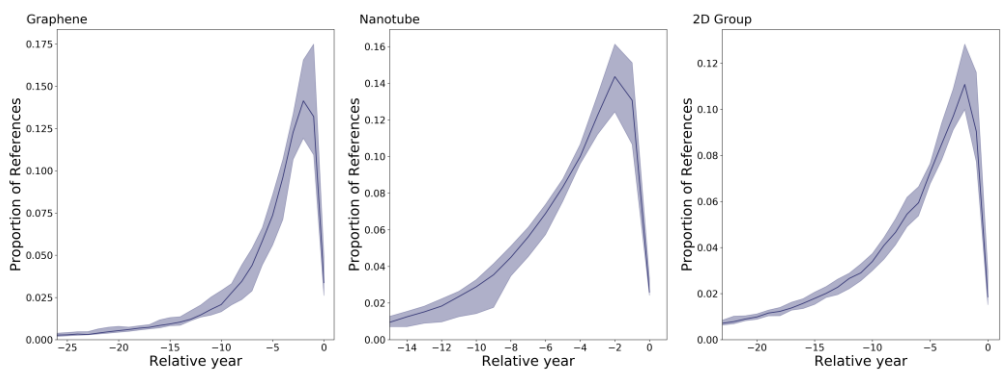

Figure 20: Citation profiles of (left) G-S, (middle) NT-S, (right) (2D+TMO+TMD)-S collections. In this figure, the median is shown as solid line, and the shaded region is between 25% and 75% percentile.

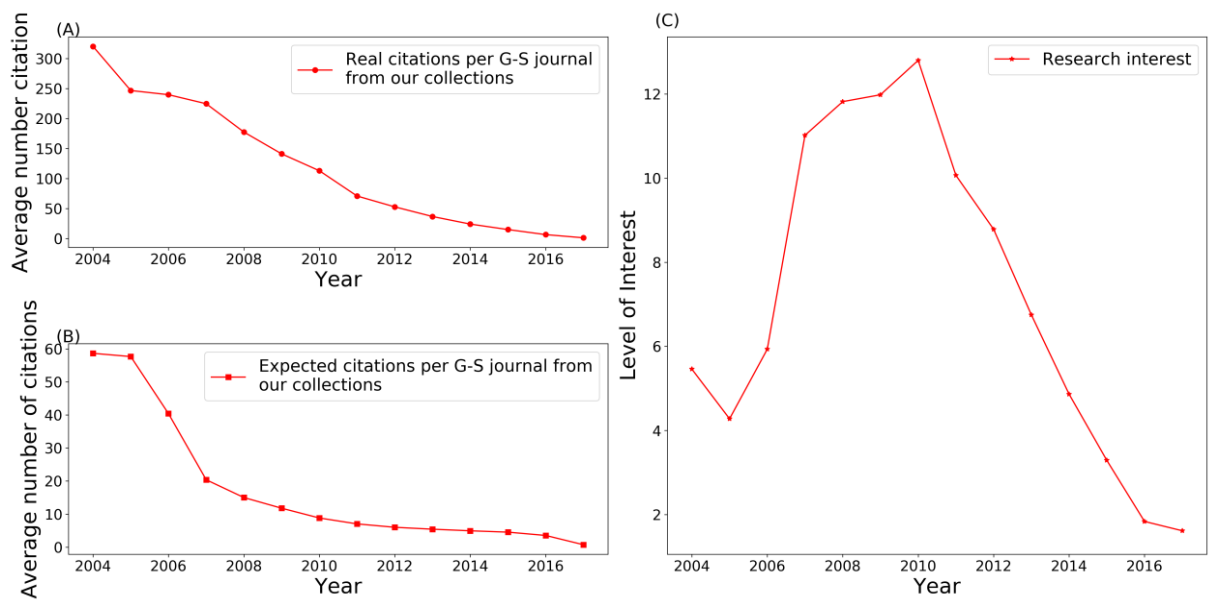

Figure 21: (A) The average number of citations per G-S journal paper between 2004 and 2017, summing contributions from the G-S, NT-S, (2D+TMO+TMD)-S collections. (B) The expected average number of citations per G-S journal paper between 2004 and 2017, based on the numbers of G-S, NT-S, (2D+TMO+TMD)-S journal papers published each year, the citation profiles shown in Figure 20 and proportions of references that are G-S journal papers shown in Figure 19. (C) The ratio between the actual and expected average number of citations per G-S journal paper. This ratio represents the level of interest in graphene science in each of years between 2004 and 2017.

From the three collections, we also obtain in Figure 20 their citation profiles — important ingredients for simulating the average citation curve. The shapes of all three profiles are highly similar, and they all peak at 2 years prior to publication of the citing paper. In Figure 21(A), we show the actual average number of citations per G-S journal paper between 2004 and 2017. This has a strong decreasing trend, but the decrease may or may not be explainable in terms of later G-S journal papers having shorter times to accumulate citations. To check this, we computed the expected average number of citations per G-S journal paper shown in Figure 21(B), by simulating the proportions of references of G-S, NT-S, and (2D+TMO+TMD)-S journal papers by years and by collections, and combining for each year citations of the G-S collection contributed by all collections. A quick visual inspection tells us that these two curves (actual and expected) are not the same. The ratio between the actual and expected represents a quantitative measure of the level of interest scientists have in graphene over the years. The interest level rose in the early years, reached a peak in 2010, and has since been on the decline. This is consistent with our discovery of 'Golden Eras' in graphene science that we have reported earlier in this paper.

Doing the same analysis for G-T patents, we make use of the proportions of references from the various collections (Figure 22), the citation profiles for patents from the various collections (Figure 23), and the numbers of patent publications from 2004 to 2017 in the various collections. In Figure 24(A), apart from a brief dip in 2007 and 2008, the actual average number of citations per G-T patent decreases monotonically till 2017. As was the case for graphene science, this decrease may or may not be due to a decrease in interest. To discriminate between these two scenarios, we simulated the expected average number of citations per G-T patent (Figure 24(B)), and calculate the ratio of actual over expected (Figure 24(C)). From 2004 to 2007, the interest in graphene technology dropped slightly, before rising strongly again. This strong rise ended in 2012 and the interest level in graphene technology has been on the decline since. We stress here that while the actual and expected average number of citations per G-T patent are both forward looking quantities, and may both change as more citations become available, the ratio which indicates the level of technological interest is a point quantity, and is not expected to change as more citations become available.

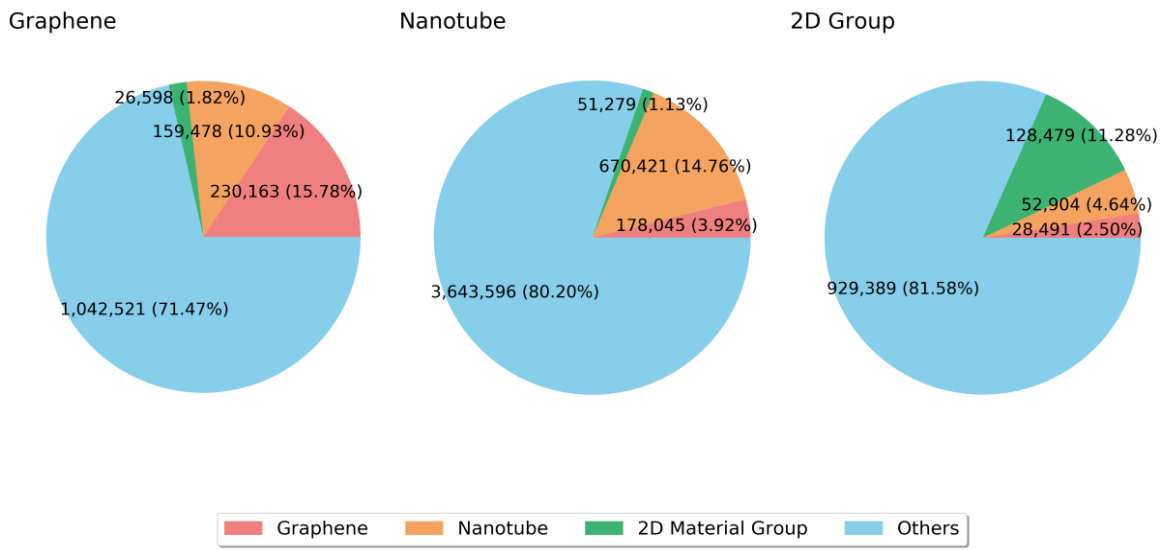

Figure 22: References of (left) G-T, (middle) NT-T, (right) (2D+TMO+TMD)-T patents sorted according to which collections they come from.

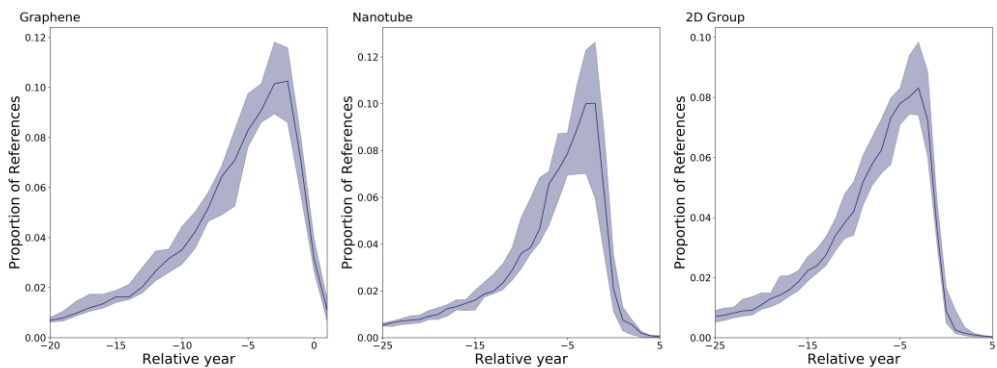

Figure 23: Citation profiles of (left) G-T, (middle) NT-T, (right) (2D+TMO+TMD)-T collections. In this figure, the median is shown as solid line, and the shaded region is between 25% and 75% percentile.

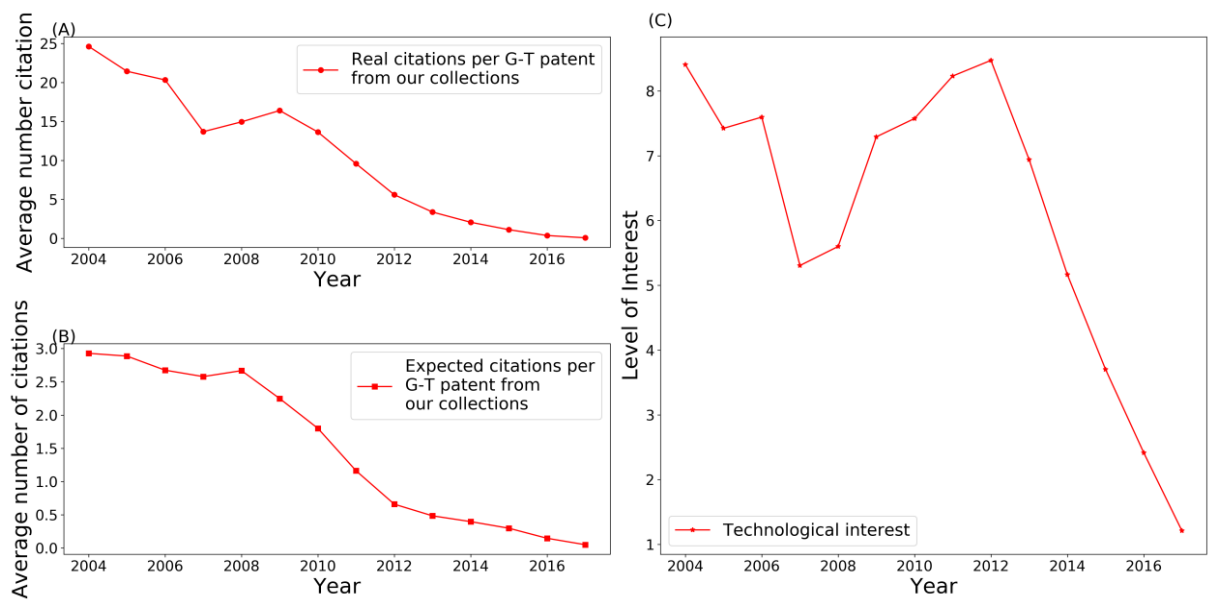

Figure 24: (A) The average number of citations per G-T patent between 2004 and 2017, summing contributions from the G-T, NT-T, (2D+TMO+TMD)-T collections. (B) The expected average number of citations per G-T patent between 2004 and 2017, based on the numbers of G-T, NT-T, (2D+TMO+TMD)-T patents published each year, the citation profiles shown in Figure 23, and proportions of references that are G-T patents shown in Figure 22. (C) The ratio between the actual and expected average number of citations per G-T patent. This ratio represents the level of interest in graphene technology in each of years between 2004 and 2017.

## 6. Conclusions

In this paper, we analyzed the collections of journal publications and patents related to graphene from three perspectives: aggregate, temporal dynamics, and geographical distribution. The journal collection is our proxy for graphene science (G-S), whereas patent collection is our proxy for graphene technology (G-T). We wished to understand graphene science and graphene technology in terms of input and output. At the aggregate level, we see G-S journal publications citing mostly G-S journal publications, although there is also significant fraction of other journal publications been cited. Most of the journal publications citing G-S journal publications are also from this same collection, meaning that G-S draws inspiration mostly from within G-S, and the impact of graphene science is limited to within graphene science, at least up to 2017. In contrast only about 16% of the patent references of G-T patents are from the G-T collection. The majority of these patent references are from outside of G-T. However, 42% of patents citing the G-T patents are themselves G-T patents. This tells us that G-T absorbs knowhow from a very broad technological field beyond G-T itself, but its impact is still somewhat limited to G-T.

From the temporal dynamics perspective, the numbers of G-S journal publications and G-T patents are increasing strongly since 2004. The number of G-S journal publications was comparable to the number of G-T patents in the beginning, but starting 2011 the yearly number of G-T patents increases more rapidly than the yearly number of G-S journal publications. By breaking down the references of G-S journal publications into the various components, we see G-S journal publications citing more G-S journal publications than other collections between 2008 and 2015. Similarly, by sorting the patents citing G-T patents into the various collections, we find that G-T patents were cited by mostly G-T patents only between 2009 and 2012. This suggests that there is a 'golden era' in G-S between 2008 and 2015, and a 'golden era' in G-T between 2009 and 2012. Evidence for these 'golden eras' can also be seen from the ratios of actual over expected average numbers of citations per publication (G-S or G-T), which shows a strong peak in 2010 and 2012 respectively. We post a caveat here: in computing the ratios of actual over expected average number of citations per publications, we have restricted ourselves to the graphene, nanotube, and 2D-material collections. However, for another collection that we may have missed to have a strong impact on the ratio computed, this other research field must demonstrate an interest in graphene that is stronger than the interest from the graphene itself. We are not aware of the existence of such field.

Finally, from the geographical distribution perspective, we see that China, US, South Korea, India, Iran, and Japan are the top six regions in graphene science, while China, US, South Korea, WPO, Japan and EPO are the top six patent offices in graphene technology. The differences between science and technology extend into their citation patterns: most patents cite other patents from same regional patent offices except for EPO and WPO, whose patents cite mostly US patents. We show that this is likely due to the languages used by the different patent offices. In contrast, G-S journal papers from various regions cited G-S journal papers from China and US, the two clear leaders in graphene science. We also tracked the scientific and technological rankings of the top ten regions over time, and found that France, an early scientific pioneer of graphene, fall out of top ten permanently in 2011, whereas regions like China, South Korea, and India rose strongly in 2006, 2008, and 2009 respectively. For graphene technology, China was the leader in graphene technology until 2000, when it was displaced by Japan and US, before regaining this leader position in 2012. But while US remained firmly behind China, Japan was displaced by South Korea in 2013.

The approach we described in this paper can also be used to test other hot research topics, like quantum technology or artificial intelligence, to see if they can stand up to this level of scrutiny.

**Acknowledgement**

This research is supported by the Singapore Ministry of Education Academic Research Fund, under the grant number MOE2017-T2-2-075. We thank Clarivate Analytics for providing us with a complimentary Derwent Innovation account to collect the patent data used in this paper.

**References**

Auranen, O., & Nieminen, M. (2010). University research funding and publication performance—An international comparison. *Research Policy*, *39*(6), 822–834. https://doi.org/10.1016/j.respol.2010.03.003
Avouris, P., & Dimitrakopoulos, C. (2012). Graphene: synthesis and applications. *Materials Today*, *15*(3), 86–97.
Carpenter, M. P., Cooper, M., & Narin, F. (1980). Linkage between basic research literature and patents. *Research Management*, *23*(2), 30–35.
Centre for Advanced 2D Materials. (n.d.). Retrieved July 26, 2019, from https://graphene.nus.edu.sg/

China No 1 in world patent applications for graphene tech - Chinadaily.com.cn. (n.d.). Retrieved July 31, 2019, from http://www.chinadaily.com.cn/cndy/2018-02/01/content_35623375.htm

Cimini, G., Gabrielli, A., & Sylos Labini, F. (2014). The Scientific Competitiveness of Nations. *PLOS ONE*, *9*(12), e113470. https://doi.org/10.1371/journal.pone.0113470

CORDIS | European Commission. (n.d.). Retrieved July 26, 2019, from https://cordis.europa.eu/

Cutler, R. S. (1987). Patents resulting from NSF's engineering program. *World Patent Information*, *9*(1), 38–42. https://doi.org/10.1016/0172-2190(87)90193-1

Derwent Innovation. (n.d.). Retrieved August 27, 2019, from Derwent website: https://clarivate.com/derwent/solutions/derwent-innovation/

Digital Trends. (n.d.). Retrieved July 26, 2019, from https://www.digitaltrends.com/cool-tech/what-is-graphene/

Ernst, H. (1997). The use of patent data for technological forecasting: the diffusion of CNC-technology in the machine tool industry. *Small Business Economics*, *9*(4), 361–381.

Funk, R. J., & Owen-Smith, J. (2017). A Dynamic Network Measure of Technological Change. *Management Science*, *63*(3), 791–817. https://doi.org/10.1287/mnsc.2015.2366

Gastner, M. T., & Newman, M. E. J. (2004). Diffusion-based method for producing density-equalizing maps. *Proceedings of the National Academy of Sciences of the United States of America*, *101*(20), 7499. https://doi.org/10.1073/pnas.0400280101

Graphene Flagship | Graphene Flagship. (n.d.). Retrieved July 26, 2019, from http://graphene-flagship.eu/

Graphene Market Analysis by Size, Price, Demand, Growth | Industry Report 2019-2025. (n.d.). Retrieved July 31, 2019, from https://www.adroitmarketresearch.com/industry-reports/graphene-market

Griliches, Z. (1990). Patent Statistics as Economic Indicators: A Survey. *Journal of Economic Literature*, *28*(4), 1661–1707. Retrieved from JSTOR.

Harhoff, D., Scherer, F. M., & Vopel, K. (2003). Citations, family size, opposition and the value of patent rights. *Research Policy*, *32*(8), 1343–1363.

Lanjouw, J. O., & Schankerman, M. A. (1998). *Patent Suits: Do They Distort Research Incentives?*

Leydesdorff, L., & Wagner, C. (2009). Macro-level indicators of the relations between research funding and research output. *Journal of Informetrics*, *3*(4), 353–362. https://doi.org/10.1016/j.joi.2009.05.005

Liu, W., Nanetti, A., & Cheong, S. A. (2017). Knowledge evolution in physics research: An analysis of bibliographic coupling networks. *PLOS ONE*, *12*(9), e0184821. https://doi.org/10.1371/journal.pone.0184821

Matutes, C., Regibeau, P., & Rockett, K. (1996). Optimal patent design and the diffusion of innovations. *The RAND Journal of Economics*, *27*(1), 60.

Morgan, R. P., Kruytbosch, C., & Kannankutty, N. (2001). Patenting and invention activity of US scientists and engineers in the academic sector: comparisons with industry. *The Journal of Technology Transfer*, *26*(1–2), 173–183.

National Science & Technology Information Service(NTIS). (n.d.). Retrieved July 26, 2019, from https://www.ntis.go.kr/en/GpIndex.do

Nordhaus, W. D. (1967). *The optimal life of a patent*. Cowles Foundation for Research in Economics, Yale University.

Novoselov, K. S., Geim, A. K., Morozov, S. V., Jiang, D., Zhang, Y., Dubonos, S. V., … Firsov, A. A. (2004). Electric field effect in atomically thin carbon films. *Science*, *306*(5696), 666–669.

NSF - National Science Foundation. (n.d.). Retrieved July 26, 2019, from https://www.nsf.gov/

Otsu, N. (1979). A threshold selection method from gray-level histograms. *IEEE Transactions on Systems, Man, and Cybernetics*, *9*(1), 62–66.

Palacios, T. (2011). Graphene electronics: thinking outside the silicon box. *Nature Nanotechnology*, *6*(8), 464.

Putnam, J. D. (1997). *The value of international patent rights*.

Rao, C. emsp14N emsp14R, Sood, A. emsp14K, Subrahmanyam, K. emsp14S, & Govindaraj, A. (2009). Graphene: the new two-dimensional nanomaterial. *Angewandte Chemie International Edition*, *48*(42), 7752–7777.

Reitzig, M. (2003). What determines patent value?: Insights from the semiconductor industry. *Research Policy*, *32*(1), 13–26.

Reitzig, M. (2004). Improving patent valuations for management purposes—validating new indicators by analyzing application rationales. *Research Policy*, *33*(6–7), 939–957.

Rossman, J., & Sanders, B. S. (1957). The patent utilization study. *Pat. Trademark & Copy. J. Res. & Ed.*, *1*, 74.

Sanders, B. S., Rossman, J., & Harris, L. J. (1958). The economic impact of patents. *Pat. Trademark & Copy. J. Res. & Ed.*, *2*, 340.

Shen, H.-W., & Barabási, A.-L. (2014). Collective credit allocation in science. *Proceedings of the National Academy of Sciences*, *111*(34), 12325–12330.

South Korea Funds Graphene Commercialization – NextBigFuture.com. (n.d.). Retrieved July 26, 2019, from https://www.nextbigfuture.com/2012/05/south-korea-funds-graphene.html

Svensson, R. (2007). Commercialization of patents and external financing during the R&D phase. *Research Policy*, *36*(7), 1052–1069.

Tanagra. (2017, October 9). Tanagra - Data Mining and Data Science Tutorials: Regression analysis in Python. Retrieved August 27, 2019, from Tanagra - Data Mining and Data Science Tutorials website: http://data-mining-tutorials.blogspot.com/2017/10/regression-analysis-in-python.html

Tanaka, H. (2019). *Fast implementation of the edit distance(Levenshtein distance): aflc/editdistance* [C++]. Retrieved from https://github.com/aflc/editdistance (Original work published 2013)

TECHNOLOGIST. (2014, July 7). Retrieved July 26, 2019, from TECHNOLOGIST | INNOVATION. EXPLAINED. website: https://www.technologist.eu/here-comes-the-graphene-revolution/

The Nobel Prize in Physics 2010. (n.d.). Retrieved July 26, 2019, from NobelPrize.org website: https://www.nobelprize.org/prizes/physics/2010/press-release

Tong, X., & Frame, J. D. (1994). Measuring national technological performance with patent claims data. *Research Policy*, *23*(2), 133–141.

Trajtenberg, M. (1990). A penny for your quotes: patent citations and the value of innovations. *The Rand Journal of Economics*, 172–187.

Web of Science. (n.d.). Retrieved from https://www.webofknowledge.com

Webster, E., & Jensen, P. H. (2011). Do patents matter for commercialization? *The Journal of Law and Economics*, *54*(2), 431–453.

Wu, Y., Welch, E. W., & Huang, W.-L. (2015). Commercialization of university inventions: Individual and institutional factors affecting licensing of university patents. *Technovation*, *36*, 12–25.

Yang, X., Yu, X., & Liu, X. (2018). Obtaining a Sustainable Competitive Advantage from Patent Information: A Patent Analysis of the Graphene Industry. *Sustainability*, *10*(12), 4800.

Zhi, Q., & Meng, T. (2016). Funding allocation, inequality, and scientific research output: an empirical study based on the life science sector of Natural Science Foundation of China. *Scientometrics*, *106*(2), 603–628. https://doi.org/10.1007/s11192-015-1773-5

Zou, L., Wang, L., Wu, Y., Ma, C., Yu, S., & Liu, X. (2018). Trends Analysis of Graphene Research and Development. *Journal of Data and Information Science*, *3*(1). Retrieved from https://content.sciendo.com/view/journals/jdis/3/1/article-p82.xml